\documentclass[]{article}
\usepackage{amssymb}
\usepackage{epsfig}
\voffset= - 0.3in
\hoffset= - 1.0in         

\catcode`\@=11
\def\marginnote#1{}

\newcount\hour
\newcount\minute
\newtoks\amorpm
\hour=\time\divide\hour by60
\minute=\time{\multiply\hour by60 \global\advance\minute by-\hour}
\edef\standardtime{{\ifnum\hour<12 \global\amorpm={am}%
        \else\global\amorpm={pm}\advance\hour by-12 \fi
        \ifnum\hour=0 \hour=12 \fi
        \number\hour:\ifnum\minute<10 0\fi\number\minute\the\amorpm}}
\edef\militarytime{\number\hour:\ifnum\minute<10 0\fi\number\minute}

\def\draftlabel#1{{\@bsphack\if@filesw {\let\thepage\relax
   \xdef\@gtempa{\write\@auxout{\string
      \newlabel{#1}{{\@currentlabel}{\thepage}}}}}\@gtempa
   \if@nobreak \ifvmode\nobreak\fi\fi\fi\@esphack}
        \gdef\@eqnlabel{#1}}
\def\@eqnlabel{}
\def\@vacuum{}
\def\draftmarginnote#1{\marginpar{\raggedright\scriptsize\tt#1}}

\def\draft{\oddsidemargin -.5truein
        \def\@oddfoot{\sl preliminary draft \hfil
        \rm\thepage\hfil\sl\today\quad\militarytime}
        \let\@evenfoot\@oddfoot \overfullrule 3pt
        \let\label=\draftlabel
        \let\marginnote=\draftmarginnote
   \def\@eqnnum{(\theequation)\rlap{\kern\marginparsep\tt\@eqnlabel}%
\global\let\@eqnlabel\@vacuum}  }

\def\appname{Appendix}
\newcounter{app}
\def\theapp{\Alph{app}}
\def\app{\par
   \addvspace{4ex}
   \@afterindentfalse
  \secdef\@app\@dapp}
\def\@app[#1]#2{\ifnum \c@secnumdepth >\m@ne
        \refstepcounter{app}
        \addcontentsline{toc}{app}{\theapp
        \hspace{1em}#1}\else
      \addcontentsline{toc}{app}{ #1}\fi
   {\parindent \z@ \raggedright
    \Large \bf \appname~\theapp .
   \Large  \bf 
    #2}\nobreak
   \vskip 4ex   \noindent
\setcounter{equation}{0}
\def\theequation{\Alph{app}.\arabic{equation}}}
\def\@dapp#1{%
{\parindent \z@ \raggedright  \bf #1}\par\nobreak}
\def\l@app#1#2{\addpenalty{\@secpenalty}%
   \addvspace{1em plus\p@}%
   \begingroup
   \@tempdima 3em
     \parindent \z@ \rightskip \@pnumwidth
     \parfillskip -\@pnumwidth
     { \bf
     \leavevmode
     #1\hfil \hbox to\@pnumwidth{\hss #2}}\par
     \nobreak
   \endgroup}

\makeatletter
\newdimen\normalarrayskip            
\newdimen\minarrayskip               
\normalarrayskip\baselineskip
\minarrayskip\jot
\newif\ifold             \oldtrue            \def\new{\oldfalse}
\def\arraymode{\ifold\relax\else\displaystyle\fi}
\def\eqnumphantom{\phantom{(\theequation)}} 
\def\@arrayskip{\ifold\baselineskip\z@\lineskip\z@
     \else
     \baselineskip\minarrayskip\lineskip1\baselineskip\fi}

\def\@arrayclassz{\ifcase \@lastchclass \@acolampacol \or
\@ampacol \or \or \or \@addamp \or
   \@acolampacol \or \@firstampfalse \@acol \fi
\edef\@preamble{\@preamble
  \ifcase \@chnum
     \hfil$\relax\arraymode\@sharp$\hfil
     \or $\relax\arraymode\@sharp$\hfil
     \or \hfil$\relax\arraymode\@sharp$\fi}}

\def\@array[#1]#2{\setbox\@arstrutbox=\hbox{\vrule
     height\arraystretch \ht\strutbox
     depth\arraystretch \dp\strutbox
width\z@}\@mkpream{#2}\edef\@preamble{\halign \noexpand\@halignto
\bgroup \tabskip\z@ \@arstrut \@preamble \tabskip\z@ \cr}%
\let\@startpbox\@@startpbox \let\@endpbox\@@endpbox
  \if #1t\vtop \else \if#1b\vbox \else \vcenter \fi\fi
  \bgroup \let\par\relax
  \let\@sharp##\let\protect\relax
  \@arrayskip\@preamble}
\def\eqnarray{\stepcounter{equation}%
              \let\@currentlabel=\theequation
              \global\@eqnswtrue
              \global\@eqcnt\z@
              \tabskip\@centering              
              \let\\=\@eqncr
              $$%
            \halign to \displaywidth  \bgroup
             \eqnumphantom \@eqnsel
      \hskip\@centering                               
    $\displaystyle  \tabskip\z@ {##}$%
    &\global\@eqcnt\@ne \hskip 2\arraycolsep
         $ \displaystyle  \arraymode{##}$\hfil
    &\global\@eqcnt\tw@ \hskip 2\arraycolsep
         $\displaystyle\tabskip\z@{##}$\hfil
         \tabskip\@centering
    &{##}\tabskip\z@\cr}
\makeatother

\newfont{\hr}{msbm10}
\newfont{\ams}{msam10}

\font\numbers=cmss12
\font\upright=cmu10 scaled\magstep1
\def\stroke{\vrule height8pt width0.4pt depth-0.1pt}
\def\topfleck{\vrule height8pt width0.5pt depth-5.9pt}
\def\botfleck{\vrule height2pt width0.5pt depth0.1pt}
\def\Zmath{\vcenter{\hbox{\numbers\rlap{\rlap{Z}\kern 0.8pt\topfleck}\kern
2.2pt
                   \rlap Z\kern 6pt\botfleck\kern 1pt}}}
\def\Qmath{\vcenter{\hbox{\upright\rlap{\rlap{Q}\kern
                   3.8pt\stroke}\phantom{Q}}}}
\def\Nmath{\vcenter{\hbox{\upright\rlap{I}\kern 1.7pt N}}}
\def\Cmath{\vcenter{\hbox{\upright\rlap{\rlap{C}\kern
                   3.8pt\stroke}\phantom{C}}}}
\def\Rmath{\vcenter{\hbox{\upright\rlap{I}\kern 1.7pt R}}}
\def\Z{\ifmmode\Zmath\else$\Zmath$\fi}
\def\Q{\ifmmode\Qmath\else$\Qmath$\fi}
\def\N{\ifmmode\Nmath\else$\Nmath$\fi}
\def\C{\ifmmode\Cmath\else$\Cmath$\fi}
\def\R{\ifmmode\Rmath\else$\Rmath$\fi}

\def\d{\partial}
\def\p{\partial}
\def\bea{\begin{eqnarray}}
\def\eea{\end{eqnarray}}

\def\beq{\begin{equation}}
\def\eeq{\end{equation}}
\def\ba{\beq\new\begin{array}{c}}
\def\ea{\end{array}\eeq}
\def\be{\ba}
\def\ee{\ea}
\def\F{{\cal F}}

\def\stackreb#1#2{\mathrel{\mathop{#2}\limits_{#1}}}
\def\Tr{{\rm Tr}}
\def\res{{\rm res}}
\def\Bf#1{\mbox{\boldmath $#1$}}

\def\bPhi{{\Bf\Phi}}
\def\bomega{{\Bf\omega}}

\def\bdelta{{\Bf\delta}}
\def\bvp{{\Bf\varpi}}

\def\bZ{{\bf Z}}

\def\bn{{\bf n}}

\def\Im{{\rm Im}}
\def\Re{{\rm Re}}

\def\half{{\textstyle{1\over2}}}
\def\ha{{1\over 2}}
\def\2{{1\over 2}}
\def\N2{${\cal N}=2$}
\def\4N{${\cal N}=4$}
\def\1N{${\cal N}=1$}
\def\1N*{${\cal N}=1^*$}
\def\x{{\sf x}}
\def\t{{\sf t}}
\def\CG{{\cal G}}

\def\f{{\sf f}}
\def\vpint{{\lefteqn{\int}{\,-}}}

\def\beq{\begin{equation}}
\def\eeq{\end{equation}}
\def\ba{\beq\new\begin{array}{c}}
\def\ea{\end{array}\eeq}
\def\be{\ba}
\def\ee{\ea}
\def\theequation{\thesection.\arabic{equation}}
\newcommand{\rf}[1]{(\ref{#1})}

\begin{document}


\begin{flushright}
FIAN/TD-18/05\\
ITEP/TH-65/05
\end{flushright}
\vspace{1.0 cm}

\renewcommand{\thefootnote}{\fnsymbol{footnote}}
\begin{center}
{\Large\bf
Matrix Models, Complex Geometry and Integrable Systems. II
\footnote{Based on lectures presented at
several schools on mathematical physics and the talks at
"Complex geometry and string theory"
and the Polivanov memorial seminar.}
}\\
\vspace{1.0 cm}
{\large A.Marshakov}\\
\vspace{0.6 cm}
{\em
Theory Department, P.N.Lebedev Physics Institute,\\
Institute of Theoretical and Experimental Physics\\ Moscow, Russia
}\\
\vspace{0.3 cm}
{e-mail:\ \ mars@lpi.ru,\ \ mars@itep.ru}
\end{center}
\begin{quotation}
\noindent
We consider certain examples of applications of the general
methods, based on geometry and integrability of matrix models,
described in \cite{MM1}. In particular, the nonlinear differential
equations, satisfied by quasiclassical tau-functions are investigated.
We also discuss a similar
quasiclassical geometric picture, arising in the context of multidimensional
supersymmetric gauge theories and the AdS/CFT correspondence.
\end{quotation}

\renewcommand{\thefootnote}{\arabic{footnote}}
\setcounter{section}{0}
\setcounter{footnote}{0}
\setcounter{equation}0

\section{Introduction}

In the first part of this paper \cite{MM1} we have discussed the properties
of the simplest gauge theories - the matrix integrals
\be
\label{mamo}
Z = \int {\rm d}\Phi e^{-{1\over\hbar}\Tr W(\Phi)}=
{1\over N!}\int \prod_{i=1}^N \left(d\phi_i e^{-{1\over\hbar}W(\phi_i)}\right)
\Delta^2(\phi)
\ee
and
\be
\label{mamocompl}
Z=\int {\rm d}\Phi {\rm d}\Phi^\dagger
\exp\left(-{1\over\hbar}V(\Phi,\Phi^\dagger)\right)=
{1\over N!}\int \prod_{i=1}^N \left(d^2z_i e^{-{1\over\hbar}V(z_i,\bar z_i)}\right)
\left|\Delta(z)\right|^2
\ee
or the so called one-matrix and two-matrix models, where by
\be
\label{V2mm}
V(\Phi, \Phi^\dagger) = \Phi^\dagger\Phi - W(\Phi;t) -
{\tilde W}({\Phi^\dagger};{\bar t})
\ee
and
\be
\label{mmpot}
W(\Phi) = \sum_{k>0} t_k \Phi^k
\ee
the corresponding gauge-invariant single-trace potentials (the last one is usually some
generic polynomial) are denoted, and
\be
\label{vdm}
\Delta (\phi) = \prod_{i<j}(\phi_i-\phi_j) = \det_{ij} \|\phi_i^{j-1}\|
\ee
stays for the Van-der-Monde determinant. We have reminded in \cite{MM1} that
matrix models \rf{mamo}, \rf{mamocompl} are effectively described in terms of
Toda integrable
systems, and demonstrated, that in planar limit $N\to\infty$ of matrices of
infinite size (or
$\hbar\to 0$ with $\hbar N = t_0 = {\rm fixed}$) these integrable systems have nice
geometric origin.

The planar limit corresponds to extracting the leading contribution to the free energy
of a gauge theory (in particular, of \rf{mamo}, \rf{mamocompl})
\be
\label{ZF}
Z_{\rm gauge} = \exp\left(F_{\rm string} \right)
\ee
which is a certain string partition function (see, e.g. \cite{Pol}), dual to
a gauge theory. In particular case of the matrix ensembles \rf{mamo} or \rf{mamocompl},
within their $1/N$-expansion \cite{1/N}
\be
\label{expan}
F_{\rm string} = \sum_{g=0}^\infty N^{2-2g}F_{g}(t,t_0)
\ee
the free energies $F_0=\F$ of the planar matrix models can be identified with
the quasiclassical tau-functions \cite{KriW}, or
prepotentials of one-dimensional complex manifolds $\Sigma$. These one dimensional
complex manifolds or complex curves have the form
\be
\label{dvc}
y^2 = W'( x)^2 + 4f( x) = \prod_{j=1}^{2n}(x-{\sf x}_j)
\ee
for the one-matrix model \rf{mamo}, and
\be\label{complcu}
z^n{\tilde z}^n + a_nz^{n+1} + {\tilde a}_n{\tilde
z}^{n+1} + \sum_{i,j\in (N.P.)_+} f_{ij}z^i{\tilde z}^j = 0
\ee
in the two-matrix case \rf{mamocompl} correspondingly; in the last formula
the sum is taken over all integer points strictly inside the
corresponding Newton polygon, see details in \cite{MM1}. To complete the
geometric formulation, these curves
should be endowed with the meromorphic generating one-forms $dS$
\be
\label{dS}
dS = \left\{\begin{array}{c}
       {i\over 4\pi}\ yd x, \ \ \ \ {\rm one-matrix\ model} \\
       {1\over 2\pi i}\ {\tilde z}dz, \ \ \ \ {\rm two-matrix\ model}
     \end{array}\right.
\ee
and the corresponding prepotentials are defined in the following way:
\be
\label{prep}
{\bf S} = \oint_{\bf A} dS,
\\
{\d\F\over\d {\bf S}} = 2\pi i\oint_{\bf B} dS
\ee
i.e. as functions of half of the periods of the generating one-form \rf{dS}
so that their gradients are given by dual periods w.r.t. the intersection form
$A_\alpha\circ B_\beta = \delta _{\alpha\beta}$. Integrability of \rf{prep} is guaranteed
by symmetricity
\be
\label{period}
{\d^2\F\over \d S_\alpha\d S_\beta} = 2\pi iT_{\alpha\beta}
\ee
of the period matrix of $\Sigma$, which is a particular case of Riemann bilinear
relations.

In the case of matrix models and their curves \rf{dvc}, \rf{complcu} the period variables
\rf{prep} have meaning of the fractions of eigenvalues, located at particular support. The
total number of eigenvalues
\be
\label{t0res}
t_0 = {1\over 2\pi i}\res_{P_+} dS = - {1\over 2\pi i}\res_{P_-} dS
\\
{\d\F\over\d t_0} = 4\pi i\int_{P_-}^{P_+} dS
\ee
and the parameters of potentials \rf{mmpot}, \rf{V2mm}
\be
\label{tP}
t_k = {1\over 2\pi ik}\ \res_{P_0} \xi^{-k}dS,\ \ \ k>0
\\
{\d\F\over \d t_k} = {1\over 2\pi i}\ \res_{P_0} \xi^{k}dS,\ \ \ k>0
\ee
are associated with the generalized periods or residues of generating one-forms;
in \rf{tP} $\xi$ is an {\em inverse} local co-ordinate at the marked point $P_0$:
$\xi(P_0)=\infty$. Analogously to \rf{period} the consistency condition for
\rf{tP} is ensured by the following formula for the second derivatives
\be
\label{sysi}
{\d^2\F\over \d t_n\d t_k} = {1\over 2\pi i}\ \res_{P_0} (\xi^k d\Omega_n)
\ee
while the third derivatives are given by the residue formula
\be
\label{residue}
{\d^3 \F\over \d T_I\d T_J\d T_K} =
{1\over 2\pi i}\ \res_{dx=0}\left(dH_IdH_JdH_K\over dx dy\right)
= {1\over 2\pi i}\sum_{x_a}\ \res_{x_a}
\left({\phi_I\phi_J\phi_K\over {dx /dy}}dy\right)
\ee
where the set of one-forms $\{ dH_I\} = \{ d\omega_\alpha, d\Omega_0, d\Omega_k\}$
corresponds to the set of parameters $\{ T_I\} = \{ S_\alpha, t_0, t_k\}$ by
\be
\label{shol}
{\d dS\over \d S_\alpha} = d\omega_\alpha, \ \ \ \alpha=1,\dots,g
\ee
together with
\be
\label{s3}
{\d dS\over \d t_0} = d\Omega_0
\ee
and
\be
\label{smer}
{\d dS\over \d t_k} = d\Omega_k, \ \ \ k\geq 1
\ee
In the second line of (\ref{residue}), for convenience in what follows,
we have introduced the meromorphic functions
\be
\label{phi}
\phi_I(x) = {dH_I\over dy} 
\ee
As functions of their parameters, being (extended) moduli of complex manifolds
\rf{dvc}, \rf{complcu} the quasiclassical tau function or prepotential
is defined only locally, but in the way consistent with the duality
transformations: $A \leftrightarrow B$, in the sense of the corresponding
Legendre transform
\be
\label{duality}
\F \leftrightarrow \F + \sum_\alpha S_\alpha\Pi_\alpha
\ee
For different choices of the basis cycles
the functions $\F$ differ merely by notation, and we will not distinguish between
different functions below when studying equations,
satisfied by quasiclassical tau-functions of matrix models.

In the next section we start with studying nonlinear differential equations, satisfied
by prepotentials or quasiclassical tau functions, which contain the dispersionless
Hirota equations and their higher genus analogs and the associativity or
Witten-Dijkgraaf-Verlinde-Verlinde (WDVV) equations.
Then we turn to various physical examples of the gauge/string duality \rf{ZF} where
the considered in \cite{MM1} methods of "geometric integrability" can be applied. We
start with the well-known topic of duality between matrix models and two-dimensional
gravity, formulated along these lines in terms of dispersionless hierarchy and then
turn to the geometric picture of the AdS/CFT correspondence.

\setcounter{equation}0
\section{Dispersionless Hirota equations
\label{ss:Hirota}}

In this section we consider the differential equations, satisfied by quasiclassical
tau-function. We first discuss in details the rational case, where $\Sigma$ is Riemann
sphere $\mathbb{P}^1$ and it is possible to write easily the explicit form of
differential equations
for the quasiclassical tau-function, appearing to be the dispersionless limit of
the Kadomtsev-Petviashvili (KP) and Toda hierarchy Hirota equations.
Finally in this section, we discuss the analogs of dispersionless
Hirota equations for the curve $\Sigma$ of higher genus and the WDVV equations.

\subsection{Geometry of dispersionless hierarchies
\label{ss:dtoda}}

Consider a generating function: the Abelian integral $S = \int^P ydx$,
corresponding to the generating form \rf{dS} of the one-matrix model. From the
formulas \rf{tP} one finds that
\be
\label{yx}
y \stackreb{x\to\infty}{=}  \sum\left( kt_k x^{k-1} + {\d \F\over\d t_k}{1\over
x^{k+1}}\right)
\ee
is expansion in local co-ordinate at $x(P)\to\infty$.
Its time derivatives (cf. with \rf{smer})
\be
\label{omkp}
\Omega_k = {\d S\over\d t_k} = x^k - \sum_j
{\d^2 \F\over\d t_k\d t_j}{1\over jx^j}
\ee
form a basis of meromorphic functions with poles at the point
$x=\infty$. Suppose now,
that these functions are globally defined on some curve $\Sigma$ and do not have
other singularities, except for those at $x=\infty$ (the case of
a single singularity exactly corresponds to the KP hierarchy).
If $\Sigma$ is sphere (or one considers the dispersionless limit of KP hierarchy),
this is enough
to define all $\Omega_k$ in terms of the "uniformizing" function
$\Omega_1=\lambda\in\mathbb{C}$.
Indeed, the set of its powers $\lambda^k$ has the same singularities as the set
of functions (\ref{omkp}), i.e. these two are related by simple
linear transformation
\be
\label{pomeg}
\Omega_1 = \lambda
\\
\Omega_2 = \lambda^2 + 2{\d^2 \F\over\d t_1^2}
\\
\Omega_3 = \lambda^3 + 3{\d^2 \F\over\d t_1^2}\lambda +
{3\over 2}{\d^2 \F\over\d t_1\d t_2}
\\
\dots
\ee
These equalities follow from the comparison of the singular at $x=\infty$
part of their expansions in $x$, following from (\ref{omkp}). Comparing
the negative
"tails" of the expansion in $x$ of both sides of eq.~(\ref{pomeg}) expresses
the derivatives
${\d^2 \F\over\d t_k\d t_l}$ (of $\Omega_k$ in the l.h.s.) in terms of only
those with $k=1$ (of $\lambda=\Omega_1$ in the r.h.s.). These relations are called
the dispersionless KP, or the dKP Hirota equations, and correspond to geometry
of rational curve with the only marked point (or when the other marked points
are simply forgotten), and they can be encoded into a "generating form"
\be
      \label{eq:7}
      (x_1 - x_2) \left (1- e^{D(x_1)D(x_2) {\cal F}}\right ) =
\left (\strut D(x_1) - D(x_2)\right )  \d_{t_1}{\cal F}
\ee
where the operators
\be\label{Dx}
D(x)=\sum_{k\geq 1}\frac{x^{-k}}{k}\p_{t_k}
\ee
are introduced.

In the case of dispersionless Toda hierarchy one should consider sphere with
two marked points and corresponding local co-ordinates $z=\infty$ and $\tilde z=\infty$
correspondingly\footnote{We follow
here the conventions of the two-matrix model case \rf{mamocompl},
corresponding to the independent choice of parameters in the vicinity of these two points.}.
Then
\be
\label{yxxy}
\tilde z \stackreb{z\to\infty}{=}  {t_0\over z} + \sum\left( kt_k z^{k-1} +
{\d \F\over\d t_k}{1\over
z^{k+1}}\right)
\\
z \stackreb{\tilde z\to\infty}{=}  {t_0\over \tilde z} + \sum\left( k{\tilde t}_k
\tilde z^{k-1} + {\d \F\over\d {\tilde t}_k}{1\over \tilde z^{k+1}}\right)
\ee
Considering now two generating functions $S = \int^P \tilde zdz$ and ${\tilde S} =
\int^P zd\tilde z$ and, like in \rf{omkp}, their time-derivatives one gets
\be
\label{omto}
\Omega_k = {\d S\over\d t_k} = z^k - \sum_j
{\d^2 \F\over\d t_k\d t_j}{1\over jz^j}, \ \ \ \ k>0
\\
{\tilde\Omega}_k = {\d \tilde S\over\d \tilde t_k} = \tilde z^k - \sum_j
{\d^2 \F\over\d \tilde t_k\d \tilde t_j}{1\over j\tilde z^j}, \ \ \ \ k>0
\ee
together with
\be
\label{omzero}
\Omega_0 = {\d S\over\d t_0} = \log z - \sum_j
{\d^2 \F\over\d t_0\d t_j}{1\over jz^j}
\\
\tilde\Omega_0 = {\d \tilde S\over\d t_0} = \log \tilde z - \sum_j
{\d^2 \F\over\d t_0\d \tilde t_j}{1\over j\tilde z^j}
\ee
Suppose again, that (\ref{omto}) and (\ref{omzero}) are now {\em globally} defined
meromorphic functions on sphere with two marked points. Then it is natural
to introduce a uniformizing function $w\in\mathbb{C}^\ast$, using the dipole differential
\be
{dw\over w} = d\Omega_0 = - d\tilde\Omega_0
\ee
since the sum of its residues should vanish. It means that
\be
\label{wz}
w \stackreb{z\to\infty}{=} ze^{- \sum_j
{\d^2 \F\over\d t_0\d t_j}{1\over jz^j}}
\\
{1\over w} \stackreb{\tilde z\to\infty}{=} \tilde ze^{- \sum_j
{\d^2 \F\over\d t_0\d \tilde t_j}{1\over j\tilde z^j}}
\ee
i.e. $w$ has a simple pole at the point, where $z=\infty$, and a simple zero at
$\tilde z=\infty$.
The Hirota equations for dispersionless Toda hierarchy come now from a simple
observation that
\be
\Omega_k = P_k(w), \ \ \
\tilde\Omega_k = \tilde P_k\left({1\over w}\right), \ \ \ k>0
\ee
for some polynomials $P_k$ and $\tilde P_k$, similar to \rf{pomeg}, some of them
were explicitly computed in \cite{C-K}. Formulas \rf{wz} are equivalent to the inverse
conformal maps for a simply-connected domain ${\sf D}$, which were
discussed in the first part of this paper \cite{MM1}.

For the two-matrix model relations \rf{sysi} can be unified into the
following expression (a combination of formulas (6.21) and (6.3) from \cite{MM1})
\be\label{Gconf1}
\log \left |
\frac{w(z)-w(z')}{w(z)\overline{w(z')} -1} \right |^2
=\log\left|{1\over z} - {1\over z'}\right|^2
+ \nabla (z)\nabla (z')\F
\ee
which implies
an infinite hierarchy of differential equations
on the function $\F$. In \rf{Gconf1} we use the operator
\be\label{D2}
\nabla (z)
=\d_{t_0} +\sum_{k\geq 1} \left (
\frac{z^{-k}}{k} \d_{t_k} +
\frac{\bar z^{-k}}{k}\d_{\bar t_k}\right )
\ee
discussed in detail in this context in \cite{MM1}, whose
holomorphic and antiholomorphic parts
\be\label{Dhol}
D(z)=\sum_{k\geq 1}\frac{z^{-k}}{k}
\p_{t_k}\,,
\;\;\;\;\;
\bar D(\bar z)=\sum_{k\geq 1}\frac{\bar z^{-k}}{k}
\p_{\bar t_k}\,,
\ee
essentially coincide with \rf{Dx}.
It is convenient to normalize $w(z)$ by $w(\infty )=\infty$ and
$\p_z w(\infty )$ to be real, i.e.
\be\label{confrad}
w(z)=\frac{z}{r}+O(1) \;\;\;
\mbox{as $z\to \infty$}
\ee
where (the real number)
$r=\lim_{z\to\infty} {dz/ dw(z)}$
is usually called the (external) conformal radius of the domain
${\sf D}$ or the eigenvalue support of the two-matrix model.
Putting $z' \to\infty$ in
(\ref{Gconf1}),
one gets
\be\label{sec3a}
\log |w(z)|^2=\log |z|^2 - \p_{t_0}\nabla (z)\F
\ee
At $z\to \infty$ this equality
yields a simple formula for the conformal radius:
\be\label{sec7}
\log r^2 = \p_{t_{0}}^2 \F
\ee
Rewriting further
(\ref{Gconf1}) in the form
\be
\log \left (
\frac{w(z)-w(z')}{w(z)\overline{w(z')} -1} \right )
-\log \left (\frac{1}{z}-\frac{1}{z'}\right )
-\left (\frac{1}{2}\p_{t_0} +D(z)\right )
\nabla (z' ) \F  =
\\
=
-\log \left (
\frac{\overline{w(z)}-
\overline{w(z')}}{w(z')\overline{w(z)} -1} \right )
+\log \left ( \frac{1}{\bar z}- \frac{1}{\bar z'}\right )
+\left (\frac{1}{2}\p_{t_0} +\bar D(\bar z)\right )
\nabla (z') \F
\ee
one gets an equality between the
holomorphic function of $z$ (in the l.h.s.) and
the antiholomorphic (in the r.h.s.) function. Therefore, both are equal to
a $z$-independent term which can be found from the
limit $z\to \infty$.
As a result, we obtain the equation
\be\label{sec6}
\log \left (
\frac{w(z)-w(z')}{w(z)- (\overline{w(z')})^{-1} } \right )
=\log \left ( 1-\frac{z'}{z}\right )
+D(z)\nabla (z') \F
\ee
which, at $z' \to \infty$, turns into the formula
for the conformal map $w(z)$:
\be\label{sec4}
\log w(z)=\log z
-\frac{1}{2}\p^{2}_{t_0}\F -\p_{t_0}D(z)\F
\ee
where we also used (\ref{sec7}).
Proceeding in a similar way, one can rearrange (\ref{sec6})
in order to write it separately
for holomorphic and antiholomorphic parts in $z'$:
\be\label{sec5}
\log \frac{w(z)-w(z')}{z-z'}\, =\,
- \,\frac{1}{2}\p_{t_0}^2\, \F +
D(z) D(z') \F
\\
-\, \log \left (1- \frac{1}{w(z)
\overline{w(z')}}\right )=
D(z)\bar D(\bar z' )\F
\ee
Writing eqs.\,(\ref{sec5})
for the pairs of points $({z_1},{z_2})$, $({z_2},{z_3})$ and $({z_3},{z_1})$ and
summing up the exponentials of the both sides of each equation
one arrives at the relation
\be\label{Hir1}
({z_1}-{z_2})e^{D({z_1})D({z_2})\F}
+({z_2}-{z_3})e^{D({z_2})D({z_3})\F}
+({z_3}-{z_1})e^{D({z_3})D({z_1})\F} =0
\ee
which is the dispersionless Hirota equation for the dKP hierarchy - a
part of the dispersionless two-dimensional Toda lattice hierarchy;
it is easy to notice, that
\rf{Hir1} is just a symmetric form of the equation \rf{eq:7}.
This equation can be regarded
as a degenerate case of the trisecant Fay
identity \cite{Fay}, see formula \rf{fay1} from Appendix~\ref{app:Fay} and
discussion of this point in next section.
It encodes the algebraic relations between the second
order derivatives
of the function $\F$. As ${z_3} \to \infty$, we get
these relations
in a more explicit but less symmetric form:
\be\label{Hir2}
1-e^{D({z_1})D({z_2} )\F}=
\frac{D({z_1})-D({z_2} )}{{z_1}-{z_2}}\,\p_{t_1}\F
\ee
which makes it clear that the totality of
second derivatives
$\F_{ij}=\p_{t_i}\p_{t_j}\F$
are expressed through the derivatives
with one of the indices put equal to unity.

More general equations of the dispersionless Toda hierarchy
are obtained in a similar way by combining eqs. (\ref{sec4}) and (\ref{sec5})
include derivatives w.r.t. $t_0$
and $\bar t_k$:
\be\label{Hir4} ({z_1}-{z_2})e^{D({z_1})D({z_2} )\F}
={z_1}e^{-\p_{t_0}D({z_1})\F}
-{z_2} e^{-\p_{t_0}D({z_2})\F}
\\
1-e^{-D(z)\bar D(\bar z )\F}=\frac{1}{z\bar z}
e^{ \p_{t_0} \nabla (z) \F}
\ee
These equations allow one to express the second derivatives
$\p_{t_m}\p_{t_n}\F$,
$\p_{t_m}\p_{\bar t_n}\F$ with $m,n\geq 1$
through the derivatives
$\p_{t_0}\p_{t_k}\F$,
$\p_{t_0}\p_{\bar t_k}\F$.
In particular, the dispersionless Toda equation,
\be\label{Toda}
\p_{t_1}\p_{\bar t_1}\F =e^{\p_{t_0}^{2}\F}
\ee
which follows from the second equation of (\ref{Hir4}) as $z \to \infty$,
expresses $\p_{t_1}\p_{\bar t_1}\F$ through $\p_{t_0}^{2}\F$.

\subsection{Generalized Hirota equations for the multisupport solution}

To derive equations for the function ${\cal F}$
in sect.~\ref{ss:dtoda}
we have used the representation (\ref{sec4})
for the conformal map $w(z)$ in terms of the derivatives of
${\cal F}$ and
eq.~(\ref{Gconf1}) relating the conformal map to the
Green function, expressed
through the second derivatives of ${\cal F}$.
In the multiply-connected case \cite{KMZ},
the strategy is basically the same, with
the analog of the conformal map
$w(z)$ (or rather of $\log w(z)$)
being the embedding of ${\sf D^c}$ into the
$g$-dimensional complex torus ${\bf Jac}$,
the Jacobian of the Schottky double $\Sigma$, see Appendix~\ref{app:Fay}.

This embedding is given, up to an overall
shift in ${\bf Jac}$, by the Abel map \rf{E1s}, where the components
$\omega_\alpha(z)=\int_{\xi_0}^{z} d\omega_{\alpha}$ of the vector
\rf{E1s} can be thought of as the holomorphic part of the
harmonic measure $\varpi_{\alpha}$, discussed in sect.~6 of \cite{MM1}.
The Abel map
is represented (see second formula in (6.57) of \cite{MM1}) through
the second order derivatives
of the function ${\cal F}$:
\be\label{E2}
\omega_\alpha(z)-\omega_\alpha(\infty)=
\int_{\infty}^{z} d\omega_{\alpha}=
-\p_{\alpha}D(z){\cal F}
\ee
and\footnote{Following \cite{KMZ},\cite{MM1} we are using in this section the time
variables $\tau_k$, which are more strictly defined in general situation, as moments
adjusted to the proper basis of functions ${\sf a}_k(z)$. In all
simple cases with finitely many nontrivial variables
they can be just identified with the Toda lattice times $t_k$, for
details see the first part of this paper \cite{MM1}.}
\be\label{E3}
2\, {\rm Re}\, \omega_\alpha(\infty)=
\varpi_{\alpha}(\infty)=-\p_{\tau_0}\p_{\alpha}{\cal F}
\ee
The Green function of the Dirichlet
boundary problem can be written in terms of the prime form
(\ref{primed}) (see Appendix~\ref{app:Fay})
on the Schottky double:
\be
\label{prime}
G(z,\zeta) = \log\left|E(z,\zeta)\over E(z, \bar\zeta)\right|
\ee
Here by $\bar\zeta$ we mean the (holomorphic) coordinate of
the ``mirror" point on the Schottky double, i.e.
the ``mirror" of $\zeta$ under the
antiholomorphic involution. The pairs of such mirror
points satisfy the condition
$\omega_\alpha(\zeta ) + \overline{\omega_\alpha( \bar \zeta)} = 0$
in the Jacobian (i.e., the sum should be
zero modulo the lattice of periods).
The prime form\footnote{Given a Riemann surface with local
coordinates $1/z$ and
$1/\bar z$ we trivialize the bundle of $-\ha$-differentials and
``redefine" the prime form $E(z,\zeta)\rightarrow
E(z,\zeta)(dz)^{1/2}(d\zeta)^{1/2}$ so that it becomes a function.
However for different coordinate patches
(the ``upper" and ``lower" sheets
of the Schottky double) one gets
different functions, see,
for example, formulas (\ref{Prime1}) and (\ref{Prime2}) below.}
is written through the Riemann theta functions
and the Abel map as follows:
\be\label{Prime1}
E(z, \zeta )=\frac{\theta_\ast
(\bomega(z)-\bomega(\zeta))}{h(z)\, h(\zeta )}
\ee
when the both points are on the upper sheet and
\be\label{Prime2}
E(z, \bar \zeta )=\frac{\theta_\ast
(\bomega(z)+\overline{\bomega(\zeta)})}{ih(z)\,
\overline{h(\zeta )}}
\ee
when $z$ is on the upper sheet and
$\bar \zeta$ is on the lower one
(for other cases we define
$E(\bar z, \bar \zeta )=\overline{E(z, \zeta )}$,
$E(\bar z, \zeta )=\overline{E(z, \bar \zeta )}$).
Here $\theta_\ast (\bomega)
\equiv \theta_{\mathbf{\delta}^\ast} (\bomega | T)$ is
the Riemann theta function (\ref{thetad})
with the period matrix $T_{\alpha \beta}=
2\pi i \, \p_{\alpha} \p_{\beta}{\cal F}$
and any odd characteristics $\mathbf{\delta}^\ast$, and
\be\label{Prime3}
h^2 (z)=-z^2 \sum_{\alpha =1}^{g}\theta_{\ast , \alpha}(0)
\p_z \omega_\alpha(z)
=z^2 \sum_{\alpha =1}^{g}\theta_{\ast , \alpha}(0)
\sum_{k\geq 1}{\sf a}_{k}'(z)\p_{\alpha}\p_{\tau_k}{\cal F}
\ee
Note that in the l.h.s. of (\ref{Prime2}) the bar
means the reflection in the double while
in the r.h.s. the bar means complex conjugation,
the notation is consistent since the local coordinate
in the lower sheet is just the complex conjugate one.
However, one should remember that $E(z, \bar \zeta)$
is {\em not} obtained from (\ref{Prime1}) by a simple
substitution of the complex conjugated argument, and
on different sheets so defined
prime ``forms" $E$ are represented by different
functions.
In our normalization (\ref{Prime2})
$iE(z, \bar z)$ is real (see also Appendix~\ref{app:Fay}) and
\be
\lim_{\zeta \to z}\frac{E(z,\zeta )}{z^{-1}-\zeta^{-1}} =1
\ee
in particular, $\lim_{z\to \infty} zE(z, \infty )=1$.

In (\ref{prime}), the $h$-functions in the
prime forms cancel, so the analog of
(\ref{Gconf1}) reads
\be
\label{gfm}
\log\left|\theta_\ast (\bomega(z)-\bomega(\zeta))
\over \theta_\ast (\bomega(z)+
\overline{\bomega(\zeta)})\right|^2=
\log\left|{1\over z} - {1\over \zeta }\right|^2
+\nabla(z)\nabla(\zeta ){\cal F}
\ee
This equation already explains the claim made in
the beginning of this section.
Indeed, the r.h.s. is the generating function for the
derivatives ${\cal F}_{ik}$ while the l.h.s. is expressed through
derivatives of the form ${\cal F}_{\alpha k}$
and ${\cal F}_{\alpha \beta}$
only, and the expansion in powers of $z, \zeta$ allows one
to express the former through the latter.

The analogs of eqs.\,(\ref{sec3a}), (\ref{sec7})
are
\be\label{gfm1}
\log\left|\theta_\ast (\bomega(z)-\bomega(\infty))
\over \theta_\ast (\bomega(z)+\overline{\bomega(\infty )})\right|^2 =
-\log|z|^2 +\p_{\tau_0} \nabla (z){\cal F}
\ee
and
\be\label{gfm2}
\log \left |\frac{h^{2}(\infty)}{\theta_{\ast}
(\bvp (\infty ))}\right |^2 =\p_{\tau_0}^{2}{\cal F}
\ee
respectively, here $\bvp (z)\equiv 2\, {\rm Re}\,\bomega (z)=
(\varpi_1 (z), \ldots , \varpi_g (z))$ and
\be
h^2 (\infty)=\lim_{z\to \infty}z\, \theta_{\ast}
\left ( \int_{\infty}^{z}d\bomega \right )=
-\sum_{\alpha =1}^{g}\theta_{\ast , \alpha}(0)
\p_{\alpha}\p_{\tau_1}{\cal F}
\ee
A simple check shows that the
l.h.s. of (\ref{gfm2}) can be written as
$-2\log (iE (\infty , \bar \infty ))$.
As is seen from the expansion
$G(z, \infty )=-\log |z| -
\log (i E(\infty , \bar \infty )) + O(z^{-1}) $ as
$z\to \infty$, $(iE (\infty , \bar \infty ))^{-1}$ is
a natural analog of the conformal radius, and
(\ref{gfm2}) indeed turns to (\ref{sec7}) in the
simply-connected case.
However, now it provides
a nontrivial relation on ${\cal F}_{\alpha\beta}$'s and
$ {\cal F}_{\alpha i}$'s:
\be
\label{corad}
\left (\sum_{\alpha}\theta_{\ast , \alpha}
\p_{\alpha}\p_{\tau_1}{\cal F}\right )
\left (\sum_{\beta}\theta_{\ast , \beta}
\p_{\beta}\p_{\bar \tau_1}{\cal F}\right )
=\theta^{2}_{\ast}
(\bvp (\infty )) e^{\p_{\tau_0}^{2}{\cal F}}
\ee
so that the ``small phase space" is defined
modulo this relation.

Now we are going to decompose these equalities into
holomorphic and antiholomorphic parts,
the results are conveniently written in terms of
the prime form. The counterpart of (\ref{sec6}) is
\be\label{analsec6}
\log \frac{E(\zeta , z)E(\infty ,
\bar \zeta )}{E(\zeta , \infty )E(z, \bar \zeta )}=
\log \left ( 1-\frac{\zeta}{z}\right )
+D(z) \nabla (\zeta ){\cal F}
\ee
and tending $\zeta \to \infty$, one gets:
\be\label{analsec4}
\log \frac{E(z, \bar \infty )}{E(z, \infty )}=
\log z +\log E(\infty , \bar \infty )-
\p_{\tau_0}D(z){\cal F}
\ee
Separating holomorphic and antiholomorphic
parts of (\ref{analsec6}) in $\zeta$, we get
analogs of (\ref{sec5}):
\be\label{analsec5}
\log \frac{E(z, \zeta )}{E(z, \infty )E(\infty , \zeta)}
=\log (z-\zeta) +D(z)D(\zeta){\cal F}
\ee
\be\label{anal511}
-\log \frac{E(z, \bar \zeta )
E(\infty , \bar \infty )}{E(z, \bar \infty )
E(\infty , \bar \zeta)} =D(z)\bar D(\bar \zeta){\cal F}
\ee
Combining these equalities, one is able to
obtain the following
representations of the prime form itself
\be\label{E10}
E(z, \zeta )=(z^{-1}-\zeta^{-1})
e^{-\frac{1}{2}(D(z)-D(\zeta ))^2 {\cal F}}
\\
iE(z, \bar \zeta )=
e^{-\frac{1}{2}(\p_{\tau_0} +D(z)+ \bar D(\bar \zeta ))^2 {\cal F}}
\ee
together with the nice formula
\be\label{E12}
iE (z, \bar z )=
e^{-\frac{1}{2}\nabla^2 (z) {\cal F}}
\ee

For higher genus Riemann surfaces there are no
simple universal relations connecting values of the
prime forms at different points, which,
via (\ref{E10}), could be used
to generate equations on ${\cal F}$.
The best available relation \cite{Fay} is the
celebrated Fay identity (\ref{fay1}). Although
it contains not only prime forms but Riemann theta
functions themselves, it is really a source
of closed equations on the quasiclassical tau-function
${\cal F}$, since all the ingredients
are in fact representable in terms of
second order derivatives of ${\cal F}$ in different variables.

An analog of the KP version of the Hirota equation (\ref{Hir1})
for the function ${\cal F}$
can be obtained
by plugging eqs.\,(\ref{E2}) and
(\ref{E10}) into
the Fay identity (\ref{fay1}). As a result, one obtains
a closed equation which contains second order derivatives
of the ${\cal F}$ only (recall that the period matrix
in the theta-functions is the matrix
of the derivatives ${\cal F}_{\alpha \beta}$).
A few equivalent forms of this equation are available:
first, shifting
$\bZ\rightarrow\bZ-\bomega_3+\bomega_4$ in (\ref{fay1}) and
putting $z_4 =\infty$, one gets the relation
\be
\label{HirKPsym}
({z_1}-{z_2})e^{D({z_1})D({z_2}){\cal F}}\
\theta \left (\int_{\infty}^{{z_1}}d\bomega + \int_{\infty}^{{z_2}}d\bomega -\bZ \right )
\ \theta \left (\int_{\infty}^{{z_3}}d\bomega -\bZ \right )\, +
\\ +  \,
({z_2}-{z_3})e^{D({z_2})D({z_3}){\cal F}}\
\theta \left (\int_{\infty}^{{z_2}}d\bomega + \int_{\infty}^{{z_3}}d\bomega -\bZ \right )
\ \theta \left (\int_{\infty}^{{z_1}}d\bomega -\bZ \right )\, +
\\ +\,
({z_3}-{z_1})e^{D({z_3})D({z_1}){\cal F}}\
\theta \left (\int_{\infty}^{{z_3}}d\bomega + \int_{\infty}^{{z_1}}d\bomega -\bZ \right )
\ \theta \left (\int_{\infty}^{{z_2}}d\bomega -\bZ \right )\, =0
\ee
where the vector $\bZ$ is arbitrary, in particular it can be chosen $\bZ=0$.
We see that (\ref{Hir1}) gets ``dressed'' by the theta-factors, and
each theta-factor is expressed through ${\cal F}$, for example
\be
\theta \left (\int_{\infty}^{z}d\bomega  \right )=
\sum_{n_{\alpha}\in \bZ}\exp \left (
 -2\pi^2 \sum_{\alpha \beta}n_{\alpha}n_{\beta}
\p_{\alpha \beta}^{2}{\cal F} -
2\pi i \sum_{\alpha}n_{\alpha}
\p_{\alpha}D(z){\cal F}\right )
\ee
Another form of this equation, obtained from
(\ref{HirKPsym}) for a particular choice of $\bZ$, reads
\be\label{HirKPsym1}
({z_1} -  {z_2}){z_3}^{-1}e^{\frac{1}{2}(D({z_1})+D({z_2}))^2 {\cal F}}
h({z_3})
\theta_{\ast}
\left (\int_{\infty}^{{z_1}}d\bomega  +  \int_{\infty}^{{z_2}}d\bomega \right ) +
[\mbox{cyclic per-s of ${z_1},{z_2},{z_3}$}] =0
\ee
Taking the limit
${z_3}\to \infty$ in (\ref{HirKPsym}), one gets
an analog of (\ref{Hir2}):
\be
\label{HirKP2}
1 \, - \,
{\theta (\int_{\infty}^{{z_1}}d\bomega +\int_{\infty}^{{z_2}}
d\bomega-\bZ)\ \ \theta (\bZ)\over
\theta (\int_{\infty}^{{z_1}}d\bomega -\bZ)\
\theta ( \int_{\infty}^{{z_2}}d\bomega-\bZ)}
\, e^{D({z_1})D({z_2}){\cal F}} \, =
\\
= \, \frac{D({z_1})-D({z_2})}{{z_1}-{z_2}}\d_{\tau_1}{\cal F}
+\frac{1}{{z_1}-{z_2}}\, \sum_{\alpha=1}^g
\frac{\p}{\p Z_{\alpha}}
\log {\theta (\int_{\infty}^{{z_1}} d\bomega
-\bZ)\over \theta (\int_{\infty}^{{z_2}}\bomega -\bZ)}
\, \p_{\alpha}  \p_{\tau_1}{\cal F}
\ee
which also follows from another Fay identity (\ref{fay2}).

Equations on ${\cal F}$ with $\bar \tau_k$-derivatives follow
from the general Fay identity (\ref{fay1}) with
some points on the lower sheet.
Besides, many other equations can be derived as various
combinations and specializations of the ones mentioned
above. Altogether, they form an infinite hierarchy of
consistent differential equations of a very complicated
structure which deserves further investigation.
The functions ${\cal F}$ corresponding to different choices
of independent variables (i.e., to different bases in
homology cycles on the Schottky double) provide different
solutions to this hierarchy.

Let us show how the simplest equation of the hierarchy,
the dispersionless Toda equation (\ref{Toda}),
is modified in the multiply-connected case.
Applying $\p_z \p_{\bar \zeta}$
to both sides of (\ref{anal511}) and setting
$\zeta =z$, we get:
\be
(\p D(z))(\bar \p \bar D(\bar z)){\cal F} =
-\p_z \p_{\bar z}\log E(z, \bar z)
\ee
where $\p D(z)$ is the $z$-derivative of the operator
$D(z)$: $\p D(z)=\sum_k {\sf a}'_{k}(z)\p_{\tau_k}$.
To transform the r.h.s.,
we use the identity (\ref{fay3})
from Appendix~\ref{app:Fay} and
specialize it to the particular local parameters
on the two sheets:
\be
|z|^4 \p_z \p_{\bar z}\log E(z, \bar z)=
\frac{\theta (\bvp (z) +\bZ )
\theta (\bvp (z)-\bZ )}{\theta^2 (\bZ )
E^2 (z,\bar z)} +
|z|^4 \sum_{\alpha, \beta}(\log
\theta ({\bf Z}))_{, \, \alpha \beta}
\, \p_z \omega_\alpha(z)\p_{\bar z}
\overline{\omega_{\beta}(z)}
\ee
where $\theta ({\bf Z}))_{, \, \alpha \beta}$ is defined in \rf{A12}.
Tending $z$ to $\infty$,
we obtain a family of equations
(parameterized by an arbitrary vector $\bZ$) which
generalize the dispesrionless Toda equation for the
quasiclassical tau-function
\be\label{Todagen}
\p_{\tau_1} \p_{\bar \tau_1} {\cal F}
=\frac{\theta (\bvp (\infty )  + \bZ )
\theta (\bvp (\infty ) -  \bZ )}{\theta^2 (\bZ )}
\, e^{\p_{\tau_0}^{2}{\cal F}}\,
-   \sum_{\alpha , \beta =1}^{g}
  (\log \theta (\bZ ))_{, \, \alpha \beta}
\, (\p_{\alpha} \p_{\tau_1}{\cal F})
(\p_{\beta} \p_{\bar \tau_1}{\cal F})
\ee
(here we used the $z\to \infty$ limits
of (\ref{E2}) and (\ref{E12})).
The following two equations correspond to
special choices of the vector $\bZ$:
\be\label{Toda2}
\p_{\tau_1} \p_{\bar \tau_1} {\cal F}
+ \sum_{\alpha , \beta =1}^{g}
(\log \theta (0))_{, \, \alpha \beta}
\, (\p_{\alpha} \p_{\tau_1}{\cal F})
(\p_{\beta} \p_{\bar \tau_1}{\cal F})
=\frac{\theta^2 (\bvp (\infty ))}{\theta^2 (0)}
\ e^{\p_{\tau_0}^{2}{\cal F}}\,
\ee
and
\be
\label{Toda1}
\p_{\tau_1} \p_{\bar \tau_1} {\cal F}
=- \sum_{\alpha,\beta =1}^{g}
\left[\log\theta_{\ast}(\bvp (\infty))\right]_{,\alpha\beta}
(\p_{\alpha} \p_{\tau_1}{\cal F})
(\p_{\beta} \p_{\bar \tau_1}{\cal F})
\ee
Finally, let us specify the equation (\ref{Todagen})
for the genus $g=1$ case, in this case there is only one extra
variable $S$, which is either $S_1$ or $\Pi_1$.
The Riemann theta-function
$\theta (\bvp (\infty ) +\bZ )$
is then replaced by the Jacobi theta-function
$\vartheta \left (\left. \p_{S}\p_{\tau_0}{\cal F}
 -Z \right |T \right )
\equiv \vartheta_3 \left (\left. \p_{S}\p_{\tau_0}{\cal F}
 -Z \right |T \right )$,
where the elliptic modular parameter is
$T =2\pi i \, \p_S^2 {\cal F}$,
and the vector $Z\equiv {\bf Z}$ has only one component.
The equation has the form:
\be\label{Todagen1}
\p_{\tau_1} \p_{\bar \tau_1} {\cal F}
=\frac{\vartheta_3 \left (\left. \p_{S}\p_{\tau_0}{\cal F}
 + Z \right |  2\pi i  \p_S^2 {\cal F} \right )
\vartheta_3 \left (\left. \p_{S}\p_{\tau_0}{\cal F}
  -  Z \right |2\pi i \, \p_S^2 {\cal F}\right )}{\vartheta_3^2
\left (\left. Z \right |2\pi i \, \p_S^2 {\cal F} \right )}
 e^{\p_{\tau_0}^{2}{\cal F}} -
\\
- \p_{Z}^{2}\log \vartheta_3
\left (\left. Z \right |2\pi i  \p_S^2 {\cal F} \right )
 (\p_S \p_{\tau_1}{\cal F})
(\p_S \p_{\bar \tau_1}{\cal F})
\ee
Note also that equation (\ref{corad}) acquires the form
\be\label{corad1}
(\p_S \p_{\tau_1}{\cal F})
(\p_S \p_{\bar \tau_1}{\cal F})=
\left (
\frac{\vartheta_1 \left (\left. \p_{S}\p_{\tau_0}{\cal F}
\right |  2\pi i \,
\p_S^2 {\cal F} \right )}{\vartheta_{1}'
\left (\left. 0 \right |  2\pi i \, \p_S^2 {\cal F} \right )}
\right )^2
e^{\p_{\tau_0}^{2}{\cal F}}
\ee
where $\vartheta_\ast\equiv\vartheta_1$ is the only
odd Jacobi theta-function. Combining (\ref{Todagen1}) and (\ref{corad1})
one may also write the equation
\be\label{Todagell}
\p_{\tau_1} \p_{\bar \tau_1} {\cal F}
= \left[
\frac{\vartheta_3 \left (\left. \p_{S}\p_{\tau_0}{\cal F}
  +  Z \right |  2\pi i \, \p_S^2 {\cal F} \right )
\vartheta_3 \left (\left. \p_{S}\p_{\tau_0}{\cal F}
  -  Z \right |2\pi i \, \p_S^2 {\cal F}\right )}{\vartheta_3^2
\left (\left. Z \right |2\pi i \, \p_S^2 {\cal F} \right )}
\right.
\\
\left.-
\left (
\frac{\vartheta_1 \left (\left. \p_{S}\p_{\tau_0}{\cal F}
\right |  2\pi i \,
\p_S^2 {\cal F} \right )}{\vartheta_{1}'
\left (\left. 0 \right |  2\pi i \, \p_S^2 {\cal F} \right )}
\right )^2
\p_{Z}^{2}\log \vartheta_3
\left (\left. Z \right |2\pi i \, \p_S^2 {\cal F} \right )
\right] \, e^{\p_{\tau_0}^{2}{\cal F}}
\ee
whose form literally reminds (\ref{Toda}), but
differs by the nontrivial
``coefficient" in the square brackets. In the
limit $T \to i\infty$ the theta-function $\vartheta_3$
tends to unity, and we obtain the dispersionless
Toda equation (\ref{Toda}).

\subsection{WDVV equations
\label{ss:WDVV}}
Let us finally briefly discuss the
Witten-Dijkgraaf-Verlinde-Verlinde or WDVV equations \cite{WDVV} in the
context of geometry of matrix models.
In the most general form \cite{MMM}
they can be written as system of algebraic relations
\be
\label{WDVV}
\F_I{\F}_J^{-1}\F_K = \F_K{\F}_J^{-1}\F_I, \ \ \ \ \ \ \forall\ I,J,K
\ee
for the matrices of third derivatives
\be
\label{matrF}
\|{\F}_{I}\|_{JK}=
{\d^3\F\over\d T_I\,\d T_J\,\d T_K} \equiv\F_{IJK}
\ee
of quasiclassical tau-function $\F ({\bf T})$. Have been appeared
first in the context of
topological string theories \cite{WDVV}, they were rediscovered later
on in much larger class of physical theories where the exact results
can be
expressed through a single holomorphic function of several complex variables,
like we have here for the planar matrix models.

Having the residue formula (\ref{residue}) for the quasiclassical tau-function,
the proof of the WDVV equations
(\ref{WDVV}) is reduced to solving the system of linear equations
\cite{BMRWZ,MaWDVV}, which requires only fulfilling the two conditions:
\begin{itemize}

\item The matching condition
\be
\label{matching}
\#(I)=\#(a )
\ee
and

\item nondegeneracy of the matrix built from (\ref{phi}):
\be
\label{detW}
\det_{Ia }\| \phi_{I}(x_a )\| \neq 0
\ee
\end{itemize}

Under these conditions, the structure constants
$C_{IJ}^K$ of the associative algebra
\be\label{CC}
\left(C_I\right)^K_L \left(C_J\right)^L_M = \left(C_J\right)^K_L
\left(C_I\right)^L_M
\\
(C_I)_J^K \equiv C_{IJ}^K
\ee
responsible for the
WDVV equations can be found
from the system of {\em linear equations}
\be
\label{eqc}
\phi_I(x_a )\phi_J(x_a ) =\sum_K
C^K_{IJ}\phi_K(x_a ), \ \ \ \ \ \ \ \forall\ x_a
\ee
with the solution
\be
\label{litc}
C^K_{IJ} = \sum_a
\phi_I(x_a )\phi_J(x_a )
\left(\phi_K(x_a )\right)^{-1}
\ee
To make it as general, as in \cite{MMM}, one may consider an
isomorphic associative algebra (again $\forall\ x_a $)
\be
\label{eqcxi}
\phi_I(x_a )\phi_J(x_a ) =\sum_K
C^K_{IJ}(\xi)\phi_K(x_a )\cdot\xi(x_a )
\ee
which instead of (\ref{litc}) leads to
\be
\label{litcxi}
C^K_{IJ}(\xi) = \sum_a
{\phi_I(x_a )\phi_J(x_a )\over\xi(x_a )}
\left(\phi_K(x_a )\right)^{-1}
\ee
The rest of the proof is consistency of this algebra with relation
\be
\label{feta}
{\cal F}_{IJK} = \sum_L C_{IJ}^L(\xi)\eta_{KL}(\xi)
\ee
with
\be
\label{metric}
\eta_{KL}(\xi) = \sum_A \xi_A {\cal F}_{KLA}
\ee
expressing structure constants in terms of the third derivatives of free energy
and, thus, leading immediately to the equations (\ref{WDVV}).
It is easy to see that (\ref{feta}) is satisfied
if ${\cal F}_{KLA}$ are given by residue formula (\ref{residue}), which can be also
represented as
\be
\label{resgen}
{\d^3 \F\over \d T_I\d T_J\d T_K} = {1\over 2\pi i}\sum_{x_a}
\res_{x_a}\left(dH_IdH_JdH_K\over dx dy\right) =\\=
{1\over 2\pi i}\sum_{x_a}\ \res_{x_a}
\left({\phi_I\phi_J\phi_K\over {dx /dy}}dy\right) =
\sum_{x_a}\Gamma_a^2
\phi_I(x_a)\phi_J(x_a)
\phi_K(x_a)
\ee
where $\Gamma_a =
\sqrt{\prod_{b\neq a}(x_a-x_b)}$, see sect.~4.5 of \cite{MM1}.

Indeed \cite{MaWDVV}, requiring only matching $\#(a )=\#(I)$, one gets
\be
\sum_K C_{IJ}^K(\xi)\eta_{KL}(\xi) =
\sum_{K,a ,b }
{\phi_I(x_a )\phi_J(x_a )\over\xi(x_a )}\cdot
\left(\phi_K(x_a )\right)^{-1}\cdot\phi_K(x_b )
\phi_L(x_b )\xi(x_b )\Gamma_b
\ee
and finally
\be
\sum_K C_{IJ}^K(\xi)\eta_{KL}(\xi) =
\sum_a
{\phi_I(x_a )\phi_J(x_a )\over\xi(x_a )}
\phi_L(x_a )\xi(x_a )\Gamma_a  =
\sum_a  \Gamma_a
\phi_I(x_a )\phi_J(x_a )\phi_L(x_a )
 = {\cal F}_{IJL}
\ee
Hence, for the proof of (\ref{WDVV}) one has to adjust the total number of
parameters $\{ T_I\}$ according to the condition (\ref{matching}). The number of critical
points for the one-matrix model is $\#(a ) = 2n$ since $dx=0$ in the
branching points of \rf{dvc}, i.e. at $y^2 = R(x) = 0$.
Thus, one have to take a codimension one subspace in the space of all
possible parameters of the one-matrix model, a natural choice will be to fix the eldest
coefficient of (\ref{mmpot}). Then the total number of parameters
$\#(I)$, including the periods ${\bf S}$, residue $t_0$ and the rest of
the coefficients of the potential will be $g+1+n = (n-1)+1+n = 2n$, i.e.
exactly equal to the number of critical points $\#(a )$.
In \cite{ChMMV} an explicit
check of the WDVV equations has been performed for this choice of parameters,
using the expansion of
free energy computed in \cite{CIV}. One could try to follow analogously for
the two-matrix model \rf{mamocompl} with non-hyperelliptic curve \rf{complcu},
where the number of critical points in the residue formula $\#(a )$ can be extracted
from \cite{MM1}, since our analysis here does not use any special properties
of the one-matrix case. However, the problem is that the number of critical points for
the two-matrix model $\#(a ) = 2n^2 + n -1$ (see e.g. formula (5.27) from \cite{MM1}) for
generic large $n$ exceeds the naive total number of parameters of the model.

Another important issue is that
formula (\ref{resgen}), expressing the third derivatives of quasiclassical
tau-function $\F$ through
the quantities $\phi_{I\alpha}$ determined in (\ref{phi}), suggests to
interpret ${\cal F}$ as the free energy of a certain topological
string theory, with
three propagators $\phi_{I\alpha}$ ending at the same three-vertex:
$$
\begin{picture}(90,55)(10,10)
\thicklines
\put(20,40){\line(1,0){20}}
\put(20,44){\makebox(0,0)[cb]{$I$}}
\put(51,58){\makebox(0,0)[lc]{$J$}}
\put(51,22){\makebox(0,0)[lc]{$K$}}
\put(20,40){\circle*{2}}
\put(49,58){\circle*{2}}
\put(49,22){\circle*{2}}
\put(40,40){\circle*{3}}
\put(45,38){\makebox(0,0)[cb]{$a$}}
\put(40,40){\line(1,2){9}}
\put(40,40){\line(1,-2){9}}
\end{picture}
$$
Moreover \cite{ChMMV2}, one can associate the next term $F_1$ of expansion \rf{expan}
with the one-loop diagram in this topological theory, i.e. with the
determinant $\det_{I,a}\phi_{I}(x_a)$:
$$
\begin{picture}(90,35)(10,20)
\thicklines
\put(60,40){\oval(20,20)}
\put(70,40){\circle*{3}}
\put(72,40){\makebox(0,0)[lc]{$I,a$}}
\end{picture}
$$
as
\be
\label{F1}
F_1= -\frac1{2}\log\left(\Delta^{1/3}({\sf x})\cdot \det_{i,j}\oint_{A_i}
{x^{j-1}dx\over y}\right)\sim \frac1{2}\log
\left(\det_{I,a}\phi_{I}(x_a)\right)
\ee
One can therefore conjecture an existence of a diagram technique for
calculating the higher genera free energies and generating function for the
correlators in matrix models.

\setcounter{equation}0
\section{Matrix models and (p,q)-string theory}

Let us now turn to some physical applications of the geometric picture
related to integrability of matrix models, which was presented
in \cite{MM1} and developed in the previous section.
As a first, already quite nontrivial example, we discuss the correlators
in the simplest, so called $(p,q)$ string theory models,
directly corresponding to the dual matrix model description of
two-dimensional gravity \cite{maint}.

\subsection{dKP for (p,q)-critical points
\label{ss:dkppq}}

According to existing about fifteen years and very popular hypothesis, see e.g. \cite{FKN},
the so called $(p,q)$ critical points of
two-dimensional gravity (corresponding to the generalized
minimal conformal matter \cite{BPZ}, interacting
with two-dimensional Liouville gravity \cite{Pol}) are most
efficiently described by
tau-function of the KP hierarchy, satisfying string equation (see also sect.~2.3 of
\cite{MM1}). In particular, it means that the correlators on world-sheets of
spherical topology (the only ones, partially computed by now by means of
two-dimensional conformal field theory) are governed by quasiclassical
tau-function of dKP hierarchy,
which is a very particular case of generic quasiclassical hierarchy. For each
$(p,q)$-th point one should consider a solution of the $p$-reduced dKP hierarchy,
or more strictly, its
expansion in the vicinity of nonvanishing $t_{p+q}={p\over p+q}$
and vanishing other times, perhaps except for
$t_1=x$ (the so called conformal backgrounds).

The geometric formulation of result in terms of quasiclassical hierarchy
is as follows:

\begin{itemize}
    \item For each $(p,q)$-th point take a pair of polynomials
\be
\label{polspq}
X=\lambda^p+\dots
\\
Y=\lambda^q+\dots
\ee
of degrees $p$ and $q$ respectively. It is convenient to rewrite
them as a generating differential
\be
\label{dSpq}
dS = YdX
\ee
\item
The parameters of quasiclassical hierarchy ("times") and the
corresponding "one-point functions" are determined by
the formulas \rf{tP}, where
\be
\label{locop}
\xi = X^{1\over p} = \lambda\left(1+
\dots + {X_0\over\lambda^p}\right)^{1\over p}
\ee
is the local co-ordinate at the point $\lambda(P_0)=\infty$. The additional
constraint that
$\xi^p=\xi^p_+$ or $\xi^p_-=0$ ($\xi^p$ is a polynomial in $\lambda$) means that
one deals here with particular $p$-reduction
of the dKP hierarchy.

\item
Under the scaling $X\to\Lambda^p X$, $Y\to\Lambda^q Y$,
(induced by $\lambda\to\Lambda\lambda$ and therefore
$\xi\to\Lambda\xi$), the times \rf{tP}
transform as $t_k \to\Lambda^{p+q-k}t_k$. Then from the second formula
of \rf{tP} it follows that the function $\F$ scales as $\F\to\Lambda^{2(p+q)}\F$,
or, for example, as
\be
\label{scaF}
\F \propto t_1^{\ 2{p+q\over p+q-1}}\ f(\tau_k)
\ee
where $f$ is supposed to be a scale-invariant function of corresponding
dimensionless ratios of the times $\tau_k=t_k/t_1^{p+q-m\over p+q-1}$ \rf{tP}.
In the simplest $(p,q)=(2,2K-1)$ case of dispersionless KdV one also
expects a natural scaling of the form
\be
\label{scakdv}
\F \propto \left(t_{2K-3}\right)^{K+\half}\ \f({\sf t}_l)
\ee
with ${\sf t}_l = t_{2l-1}/(t_{2K-3})^{(K-l+1)/2}$, with the distinguished
cosmological time $t_{2K-3}\propto \Lambda^4$.

\end{itemize}

Note also, that the tau-functions of $(p,q)$ and $(q,p)$ theories do not
coincide, but are related by the Legendre transform \cite{KhMa}, discussed
in detail for generic prepotentials in \cite{MM1}.

\subsection{Pure gravity: (p,q)=(2,3)}

In this case one has only two nontrivial parameters $t_1$ and $t_3$, and
the partition function can be calculated explicitly.
The times \rf{tP} are here expressed by
\be
\label{t23}
t_5={2\over 5}, \ \ \ t_3={2\over 3}Y_1-X_0, \ \ \ t_1={3\over 4}X_0^2-X_0Y_1
\ee
in terms of the coefficients of the
polynomials (the other coefficients vanish for $t_2=t_4=0$)
\be
\label{xy23}
X=\lambda^2+X_0, \ \ \ \ Y=\lambda^3+Y_1\lambda
\ee
and, though nonlinear, these quadratic equations
can be easily solved for the latter
\be
\label{coef23}
X_0={1\over 3}\sqrt{9t_3^2-12t_1}-t_3,
\ \ \ \
Y_1={1\over 2}\sqrt{9t_3^2-12t_1}
\ee
The second half of residues \rf{tP} gives
\be
\label{f123}
{\d\F\over\d t_1} = {1\over 8}X_0^3-{1\over 4}Y_1X_0^2
\\
{\d\F\over\d t_3} = -{1\over 8}Y_1X_0^3+{3\over 64}X_0^4
\ee
This results in the following explicit formula for the quasiclassical tau-function
\be
\label{Fpugr}
\F = {1\over 3240}\left(9t_3^2-12t_1\right)^{5/2}+
{1\over 4}t_3^3t_1-{1\over 4}t_3t_1^2-{3\over 40}t_3^5
\ee
At $t_3\to\infty$ (expansion at $t_1\to 0$) formula \rf{Fpugr} gives
\be
\F\ \stackreb{t_3\to\infty}{=}\ - {t_1^3\over 18t_3}\left(1 +
O\left({t_1\over t_3^2}\right)\right)
\ee
which is the partition function of the Kontsevich model \cite{Kontsevich,GKM}
(also identified with the $(2,1)$-point or
topological gravity). At $t_1\to\infty$ tau
function \rf{Fpugr} scales as $\F \propto t_1^{5/2}$ or partition function
of pure two-dimensional gravity \cite{maint}: expansion at $t_1\to\infty$ gives
\be
\F = (-3t_1)^{5/2}\left({4\over 405}-{1\over 54}{t_3^2\over t_1}+{1\over 96}
{t_3^4\over t_1^2}+O\left({t_3^6\over t_1^3}\right)\right) + \ldots
\ee
modulo analytic terms.

\subsection{The Yang-Lee model: (p,q)=(2,5)
\label{ss:25}}

The calculation of times according to \rf{tP} gives
\be
\label{txy25}
t_1 = -{5\over 8}X_0^3+{3\over 4}Y_3X_0^2-Y_1X_0
\\
t_3= {5\over 4}X_0^2-Y_3X_0+{2\over 3}Y_1
\\
t_5 = {2\over 5}Y_3-X_0
\\
t_7 = {2\over 7}
\ee
for the polynomials
\be
\label{pol25}
X= \lambda^2+X_0
\\
Y= \lambda^5+Y_3\lambda^3+Y_1\lambda
\ee
We see, that \rf{txy25} can be easily solved for
\be
\label{yj}
Y_1={3\over 2}\left(t_3+{5\over 4}X_0^2+{5\over 2}t_5X_0\right)
\\
Y_3 = {5\over 2}\left(X_0+t_5\right)
\ee
and
the string equation for $X_0$ is now
\be
\label{gpse}
t_1 = -{5\over 8}X_0^3-{3\over 2}t_3X_0
\ee
The one-point functions \rf{tP} are given by
\be
\label{1p25}
{\d\F\over \d t_1} = -{5\over 64}X_0^4-{1\over 4}Y_1X_0^2+{1\over 8}Y_3X_0^3
= -{15\over 64}X_0^4 - {3\over 8}t_3X_0^2
\\
{\d\F\over \d t_3} = -{1\over 8}Y_1X_0^3-{3\over 128}X_0^5+{3\over 64}Y_3X_0^4
= -{9\over 64}X_0^5 - {3\over 16}t_3X_0^3
\ee
where, in the r.h.s.'s, $Y_j$ obtained from solving \rf{txy25} are substituted.
From \rf{sysi}, or,
differentiating \rf{1p25} and using \rf{gpse}, one can easily obtain\footnote{
The second equality of \rf{252d} can be interpreted as dispersionless
Hirota equation with an additional KdV ($p=2$) reduction condition.}
\be
\label{252d}
{\d^2\F\over\d t_1^2} = {X_0\over 2}
\\
{\d^2\F\over\d t_1\d t_3} = {3\over 8}X_0^2 =
{3\over 2}\left({\d^2\F\over\d t_1^2}\right)^2
\\
{\d^2\F\over\d t_3\d t_3} = {3\over 8}X_0^3 =
3\left({\d^2\F\over\d t_1^2}\right)^3
\ee
However, \rf{gpse} cannot be effectively solved explicitly, though it can be
solved expanding in $t_3$. In the "naive" scaling regime \rf{scaF} for the partition
function of $(2,5)$ model one gets
\be
\label{25sca}
\F = t_1^{7/3}\ f\left({t_3\over t_1^{2/3}}\right)
\\
{\d\F\over \d t_1} = {7\over 3} t_1^{4/3}f - {2\over 3}t_3t_1^{2/3}f'
\\
{\d^2\F\over \d t_1^2} = {28\over 9} t_1^{1/3}f - 2{t_3\over t_1^{1/3}}f' +
{4\over 9}{t_3^2\over t_1}f''
\ee
Substituting the last expression (using \rf{252d}) into \rf{gpse} one gets
for the first coefficients of $f$
\be
\label{gpis}
f(\tau) = f_0 + f_1\tau + {1\over 2}f_2\tau^2 + {1\over 6}f_3\tau^3 +
{1\over 24}f_4\tau^4 + \ldots
\\
f_0 = - \frac {9}{140}5^{2/3}, \ \ f_1 = \frac {9}{50}5^{1/3}, \ \
f_3 = \frac {9}{25}5^{2/3}, \ \ f_4 = \frac {18}{25}5^{1/3}, \ \  \ldots
\ee
while $f_2$ (the coefficient at analytic term $t_1t_3^2$ of expansion of $\F$,
disappearing from the higher derivatives) remains undetermined.
The correspondent "scale-invariant" rations are, for example:
\be
\label{25rat}
{f_1f_3\over f_0f_4} = {7\over 5}, \ \ {f_1^3\over f_0^2f_3} = {98\over 125}, \ \
{f_1^4\over f_0^3f_4} = {686\over 625}, \ \ {f_0f_3^2\over f_1^2f_4} = {25\over 14},
\ \ \ldots
\ee
Note, that these scale-invariant rations are given by simple rational numbers, compare
to the coefficients \rf{gpis} themselves, which essentially
depend upon normalizations of times.
In Appendix~\ref{app:Ising} it is shown, that expansion \rf{gpis} is in fact related
to the generating function in the gravitationally dressed $(p,q)=(3,4)$ Ising model.

However, for the non-unitary case one may expect another scaling regime, where the
role of times is "exchanged". For the particular case of $(2,5)$ model this implies
that\footnote{Such scaling implies also a particular behavior
$F_A(t_1)=A^{-7/2}{\sf z}\left(t_1A^{3/2}\right)$
for the Laplace transformed partition function
$\F (t_1,t_3) = \int_0^\infty {dA\over A}e^{-t_3A}F_A(t_1)$
with "fixed area" (cf. with \cite{Zamol}).}
\be
\label{25sca1}
\F = t_3^{7/2}\ {\sf f}\left({t_1\over t_3^{3/2}}\right)
\equiv t_3^{7/2}\ {\sf f}(\t)
\\
{\d\F\over \d t_1} = t_3^2{\sf f}', \ \ \ \ \
{\d^2\F\over \d t_1^2} = t_3^{1/2}{\sf f}'',\ \ etc
\ee
and string equation \rf{gpse} turns into
\be
\label{gpsca}
{\sf t} +5(\f'')^3 + 3\f'' = 0
\ee
Solution of
\rf{gpsca} immediately gives for the higher coefficients of expansion
$\f_n \equiv \left.\f^{(n)}\right|_{\t=0}$ of the normalized
function ${\sf f}$ the explicit expressions
\be
\label{fsca}
{\sf f}_3 = -{1\over 3(1+5{\sf f}_2^2)}
\\
{\sf f}_4 = -{10{\sf f}_2\over 9(1+5{\sf f}_2^2)^3}
\\
{\sf f}_5 = -{10(-1+25{\sf f}_2^2)\over 27(1+5{\sf f}_2^2)^5}
\\
etc
\ee
through the 2-point function, which satisfies the cubic equation
\be
3{\sf f}_2 + 5\f_2^3 = 0
\ee
while the "normalization" ${\sf f}_0$ and the 1-point function ${\sf f}_1$ remain
undetermined from string equation \rf{gpsca} directly. In order to find them, one
should use the second equation of \rf{1p25} together with the second formula of
\rf{25sca1}, or
\be
7\f - 3\t\f'+9(\f'')^5 + 3(\f'')^3=0
\ee
giving rise to
\be
\f_0 = -{9\over 7}\f_2^5-{3\over 7}\f_2^3
\\
\f_1 = -{9\over 4}\f_2^3\f_3-{45\over 4}\f_2^4\f_3
\ee
In the "phase" with non-vanishing ${\sf f}_2$, this results in
\be
\label{f25dKP}
\f_0 = - {36\over 175}\sqrt{-{3\over 5}}, \ \ \
\f_1 = -{9\over 20},\ \ \
{\sf f}_2 = \sqrt{-{3\over 5}},\ \ \
\f_3 = {1\over 6},\ \ \
\f_4 = {1\over 4}\sqrt{-{5\over 3}},\ \ \
\f_5 = -{5\over 27},\ \ \ldots
\ee
Finally, one can easily compute the
"invariant ratios", which do not depend upon normalization of times and
partition function, for example
\footnote{Ratios \rf{ratsca} coincide with recently computed,
using the worldsheet Liouville theory, by A.~Belavin and Al.~Zamolodchikov,
see \cite{Zamol}.}
\be
\label{ratsca}
{{\sf f}_0{\sf f}_2\over {\sf f}_1^2} = {64\over 105},\ \ \
{{\sf f}_4{\sf f}_2\over {\sf f}_3^2} = -3,\ \ \
{{\sf f}_4{\sf f}_3\over {\sf f}_2{\sf f}_5} = -{1\over 8}, \ldots
\ee
Analogous formulas for the $(2,7)$ minimal string theory and $(3,4)$ or the Ising model
interacting with gravity are presented in Appendix~\ref{app:Ising}.

\subsection{Gurevich-Pitaevsky problem and matrix model with
logarithmic potential
\label{ss:gp}}

In this section we turn to discussion of the Gurevich-Pitaevsky (GP)
problem \cite{GP}, appeared originally when studying nonlinear waves,
and related later to string theory and non-perturbative gauge theories (see
e.g. \cite{MGP}) and to the problems of Laplacian growth \cite{Lapgro}, see
recent paper \cite{ZWGP}. Solution to the GP problem
connects different phases dispersionless solutions of
quasiclassical hierarchies, corresponding to
rational curves $\Sigma$ of spherical topology by seweing them via the solutions
given by the curves of higher genera. In particular, this resolves the singularities
of dispersionless solutions. The parameters of solutions
appear to be "modulated" in accordance with equations of quasiclassical hierarchy.

Below we consider, first, the original GP problem in
KdV hierarchy, whose geometry coincides with that of the $(2,5)$ model of the
previous section,
and then turn to particular example of two-matrix model \rf{mamocompl}
with non-polynomial
potential. The basic geometric condition for the GP problem is that the finite-gap
solutions
to nonlinear wave equations for the curves of different genera, in order to be correctly
sewed when overcoming the singularity, must have the same values of the (real parts of the)
periods of generating differential $\oint dS$ over the nontrivial cycles.

In one of two original GP setup's one considers the properties of the cubic
string equation \rf{gpse} of the $(2,5)$ model. The corresponding complex manifold
arises when one notices, that
the polynomials \rf{pol25}
satisfy the equation of (degenerate) $(2,5)$-curve
\be
\label{gpcu}
Y^2 = X^5 + (2Y_3-5X_0)X^4 + (Y_3^2-8Y_3X_0+2Y_1+10X_0^2)X^3+ \\+
(2Y_3Y_1-3Y_3^2X_0-6Y_1X_0-10X_0^3+12Y_3X_0^2)X^2+ \\ +
(Y_1^2+5X_0^4-8Y_3X_0^3+3Y_3^2X_0^2+6Y_1X_0^2-4Y_3Y_1X_0)X+ \\+
2Y_3Y_1X_0^2+2Y_3X_0^4-X_0^5-Y_3^2X_0^3-2Y_1X_0^3-Y_1^2X_0 \equiv
\\
\equiv (X-X_0)(X-\x_+)^2(X-\x_-)^2
\ee
with
\be
\label{fxpm}
\x_\pm = -{X_0\over 4} \pm {1\over 2}\sqrt{-{5\over 4}X_0^2-6t_3}
\ee
and the sum of the roots, proportional to $t_5$ from \rf{txy25}, to be taken vanishing.
If all
singularities are resolved, the curve \rf{gpcu} would have genus $g=2$, and in the case
of partial resolution of the singularities it can be presented as elliptic curve with an
extra "double point".

The critical parameters of the solution are determined by vanishing of
the periods of generating differential $dS = YdX$. In rational case they
can be easily computed, say, using \rf{pol25}, so that
\be
\label{pint25}
\int_{\lambda_i}^{\lambda_j} YdX =
\left.{2\over 7}\lambda^7+{2\over 5}Y_3\lambda^5+
{2\over 3}Y_1\lambda^3\right|_{\lambda_i}^{\lambda_j} =
\left.{2\over 7}\lambda^7+X_0\lambda^5+
\left(t_3+{5\over 4}X_0^2\right)\lambda^3\right|_{\lambda_i}^{\lambda_j}
\ee
where the critical points $\{\lambda_i\}$ are found
solving equation $Y=0$. Computing \rf{pint25} for these limits, one finds
the critical values
\be
{t_1\over (-t_3)^{3/2}} = {2\over\sqrt{5}},\ \  {3\sqrt{6}\over \sqrt{5}},
\ \  -{\sqrt{2}\over \sqrt{3}}
\ee
The ratio of last two values equals to ${-\sqrt{2}\over\sqrt{10}/27}$, found numerically
by Gurevich and Pitaevsky \cite{GP}, while the first one is just the critical value of
the equation \rf{gpse}. The other two values correspond to $\x_+=\x_-$ and another,
already nontrivial, case of vanishing of the integral $\int_{\x_-}^{\x_+} YdX = 0$.
From the point of view of scaling Yang-Lee model these critical values correspond to
the particular points in parameter space, where the stringy one-loop correction to the
two-point correlator $X_0$ vanishes.

\begin{figure}[tp]
\centerline{\epsfig{file=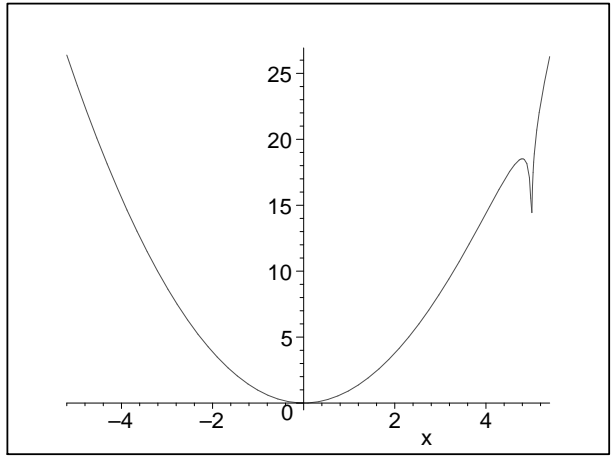,width=80mm}}
\caption{Vertical section of the matrix model potential \rf{V2mm} with
logarithmic holomorphic part \rf{jpot}, which has the cusp
at $z=a$ (here we took $a=5$ and $\nu=1$.)}
\label{fi:profile}
\end{figure}
\begin{figure}[tp]
\centerline{\epsfig{file=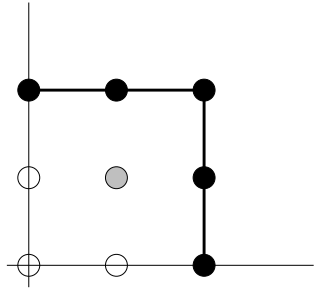,width=40mm,angle=-90}}
\caption{Newton polygon (a square!) for the curve (\ref{jcur}) of genus $g=1$.}
\label{fi:np}
\end{figure}

Let us now perform a similar procedure for the two-matrix model.
Up to now the whole geometric construction for the two matrix model in \cite{MM1}
was rather implicit, and below we discuss
an explicit (and even non-polynomial, cf. with discussion in \cite{WMiwa})
example of the potential \rf{V2mm} with the logarithmic holomorphic part
\be
\label{jpot}
W(z) = -\nu\left(\log\left(1-{z\over a}\right)+ {z\over a}\right)
\ee
taking $a$ to be real, see profile at fig.~\ref{fi:profile}. Two its extrema
satisfy the equation
\be
\label{jextreq}
z\bar z - a\bar z + {\nu\over a}z = 0
\ee
together with the complex conjugated, and are located at $z =\bar z =0$, which
is the minimum of potential and $z=\bar z = a - {\nu\over a}$, which is
the saddle point.
The corresponding equation of the complex curve \rf{complcu} for the
two-matrix model \cite{KM} reads
\be
\label{jcur}
z^2{\tilde z}^2 - \left(a-{\nu\over a}\right)\left(z^2{\tilde z} +
z{\tilde z}^2\right) + cz{\tilde z} -
\nu\left(z^2 + {\tilde z}^2\right) + g(z+\tilde z) + h \equiv
Q(z)\tilde z^2 - P(z)\tilde z - K(z) = 0
\ee
where last two terms correspond to the points
strictly inside the corresponding
Newton polygon, see fig.~\ref{fi:np}.
The curve (\ref{jcur}) is endowed with a standard two-matrix model
generating differential
\be
\label{gendiff}
\tilde z dz = \left({P\over 2Q} +
{\sqrt{R}\over 2Q}\right)dz = \left({P\over 2Q} +
{1\over 2Q}\sqrt{P^2+4QK}\right)dz
\ee
Rational degeneration of this curve (see general discussion of rational degenerations
in Appendix~A of \cite{MM1})
can be described by conformal map
\be
\label{comap}
z = rw + {u\over w-s} +v
\\
\tilde z = {r\over w} + {uw\over 1-sw} +v
\ee
where we imply all coefficients to be real; in contrast to the conformal
maps considered in Appendix~A of \cite{MM1}, the maps
\rf{comap} possess singularities not only at $w=\infty$.
In the case of rational degeneration of the curve \rf{jcur} one has
\be
\label{sora}
\sqrt{R} = \Delta(z-\xi_0)\sqrt{(z-\xi_+)(z-\xi_-)}
\\
\tilde z = {P\over 2Q} +
{1\over 2Q}\sqrt{P^2+4QK} = {P(z)\over 2Q(z)} +
{\Delta\over 2Q(z)}(z-\xi_0)\sqrt{(z-\xi_+)(z-\xi_-)}
\ee
where
\be
\label{lam}
\Delta = a + {\nu\over a}
\ee
and it corresponds to the following relations between the coefficients in
(\ref{sora}) and (\ref{comap})
\be
\label{raco}
\xi_\pm = v + rs \pm 2\sqrt{ru}
\\
\xi_0= v + {r-u\over s}
\ee
and some more complicated for the coefficients of (\ref{jcur}). Some useful technical
formulas for this case are collected in Appendix~\ref{app:Miwa}.

It is easy to see taking the residues of generating differential \rf{gendiff}, that
in addition to our common parameter
\be
\label{t0conf}
t_0 = {1\over 2\pi i}\oint_{|w|=1} {\tilde z}dz = {1\over 2\pi i}\ \res_{q_1} {\tilde z}dz +
{1\over 2\pi i}\ \res_{\infty_+} {\tilde z}dz =
{1\over 2\pi i}\ \res_{q_1} {\tilde z}dz - {1\over 2\pi i}\ \res_{q_2} {\tilde z}dz = \\ =
r^2 - {u^2\over (1-s^2)^2} = {a^2-\nu\over a^2+\nu}
\left(a^2+{\nu^2\over a^2}-c\right) - {2ga\over a^2+\nu}
\ee
it is convenient to introduce
\be
\label{delta}
\delta \equiv {1\over 2\pi i}\ \res_{q_1} {\tilde z}dz +
{1\over 2\pi i}\ \res_{q_2} {\tilde z}dz +2\nu =
a^2+{\nu^2\over a^2} -c
\ee
Using (\ref{delta}) and (\ref{t0conf}) one can rewrite equation of the curve
(\ref{jcur}) as
\be
\label{jcurd}
z^2{\tilde z}^2 - \left(a-{\nu\over a}\right)\left(z^2{\tilde z} +
z{\tilde z}^2\right) + \left(a^2+{\nu^2\over a^2}-\delta\right)z{\tilde z} -
\nu\left(z^2 + {\tilde z}^2\right) + \\ +
\left({t_0\over 2}\left(a+{\nu\over a}\right)+
{\delta\over 2}\left(a-{\nu\over a}\right)\right)(z+\tilde z) + h = 0
\ee
The last coefficient $h$ is related to the only nontrivial period of
(\ref{gendiff})
\be
\label{per}
 S = {1\over 2\pi i}\oint_A \tilde z dz
\ee
In what follows, we fix $\delta=t_0$, corresponding to
$\res_{q_2} {\tilde z}dz = {\rm const}$, then
eq.~(\ref{jcurd}) acquires the final form
\be
\label{jcurfi}
z^2{\tilde z}^2 - \left(a-{\nu\over a}\right)\left(z^2{\tilde z} +
z{\tilde z}^2\right) + \left(a^2+{\nu^2\over a^2}-t_0\right)z{\tilde z} -
\nu\left(z^2 + {\tilde z}^2\right) + t_0a(z+\tilde z) + h = 0
\ee
These formulas give rise to the following system of equations to the
parameters of conformal map (\ref{comap})
\be
\label{sysrus}
r + {u\over 1-s^2} = \Delta s
\\
\left(r + {u\over 1-s^2}\right)\left(r - {u\over 1-s^2}\right) = t_0
\\
r^2 - {ru\over s^2} = t_0-\nu
\ee
Solving (\ref{sysrus}) one gets
\be
\label{ru}
r = \ha\left(\Delta s + {t_0\over\Delta s}\right)
\\
u = {1-s^2\over 2} \left(\Delta s - {t_0\over\Delta s}\right)
\ee
and $s^2={\cal X}$ satisfies the following cubic equation
\be
\label{cueqX}
{\cal X} + {t_0^2\over 2\Delta^4}{1\over {\cal X}^2} =
\ha -{2\nu\over\Delta^2} +
{t_0\over\Delta^2}
\ee
Eq.~(\ref{cueqX}) can be solved for $t_0$
\be
\label{t0X}
t_0 = \Delta^2{\cal X}^2 \pm \Delta^2{\cal X}
\sqrt{\left({\cal X}-1-{2\sqrt{\nu}\over\Delta}\right)
\left({\cal X}-1+{2\sqrt{\nu}\over\Delta}\right)}
\ee
It is easy to see, that two solutions of (\ref{t0X}) are positive at
\begin{figure}[tp]
\centerline{\epsfig{file=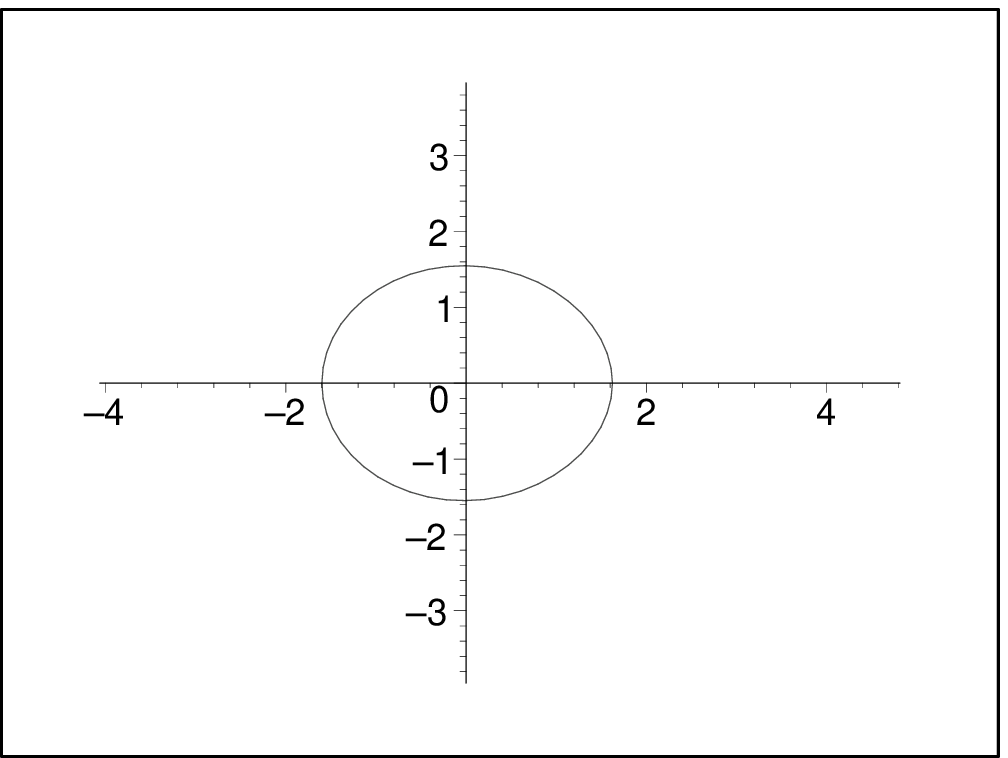,width=70mm}
\epsfig{file=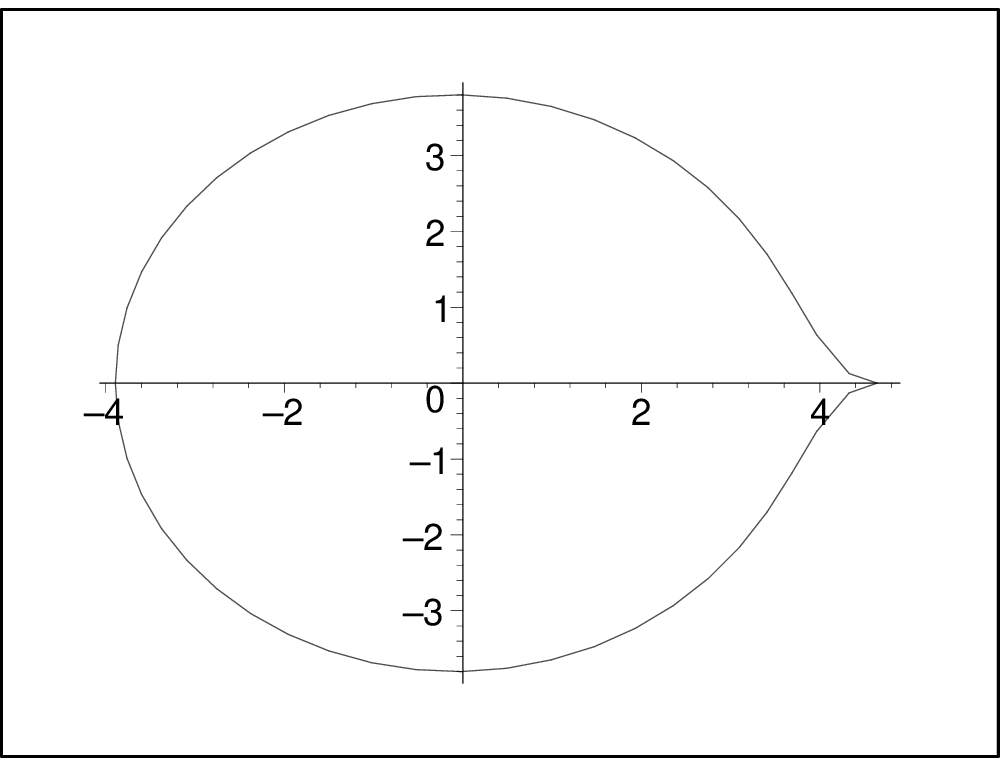,width=70mm}}
\caption{Boundary curves for the first phase. Starting like at the
left picture it develops a cusp, when approaching
$z_{\rm saddle}=a-{\nu\over a}$  (at the pictures $a=5$ and $\nu=\ha$.)}
\label{fi:loop1}
\end{figure}
\be
t_0^{(+)}: \ \ \ \ 0<{\cal X}<1-{2\sqrt{\nu}\over\Delta},\
{\cal X}>1+{2\sqrt{\nu}\over\Delta}
\\
t_0^{(-)}: \ \ \ \   {\cal X}<0,\
\ha-{2\nu\over\Delta^2}<{\cal X}<1-{2\sqrt{\nu}\over\Delta},\
{\cal X}>1+{2\sqrt{\nu}\over\Delta}
\ee
Thus, the whole picture can be presented as a growing droplet, which
is a two-dimensional horizontal surface of an "eigenvalue liquid"
of the area $t_0$ filled into the "cup" potential, see
fig.~\ref{fi:profile}.

It is easy to see, that two phases of real physical interest are:

\begin{figure}[tp]
\centerline{\epsfig{file=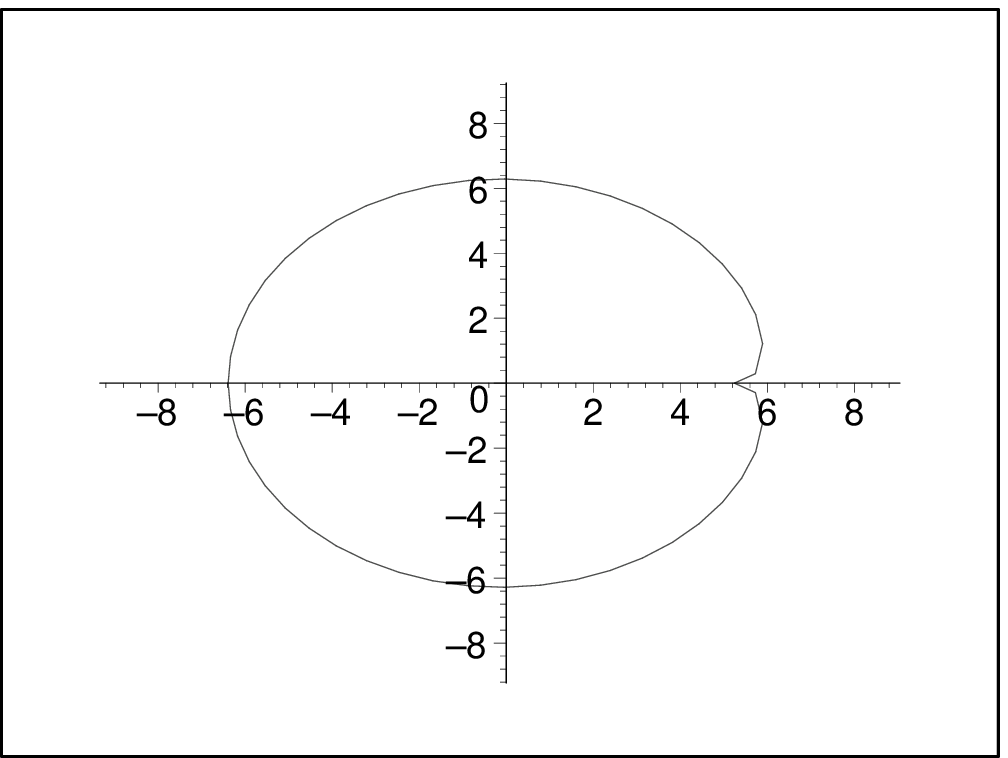,width=70mm}
\epsfig{file=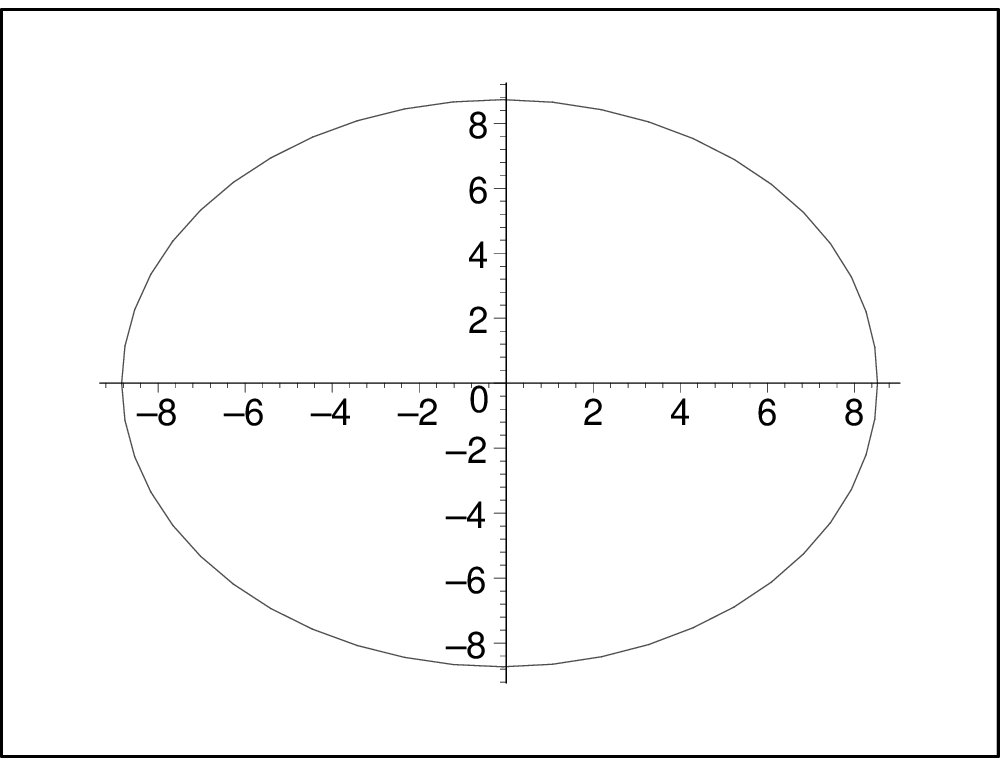,width=70mm}}
\caption{Boundary curves for the second phase. Starting from
"impurity" at $z=a$ like at the left picture, it then grows to
a smooth ellipse-like configuration like at the right picture
(again, the pictures are drawn for $a=5$ and $\nu=\ha$).}
\label{fi:loop2}
\end{figure}

\begin{itemize}
  \item Growing droplet before the "impurance": this corresponds to
$0<{\cal X}<1-{2\sqrt{\nu}\over\Delta}$ and $t_0=t_0^{(+)}$. The liquid of
eigenvalues starts to fill the minimum of two-dimensional potential at
$z=z_{\rm min}=0$ and with
growth of $t_0$ it level raises till it reaches $V(z_{\rm saddle})=
a^2-{\nu^2\over a^2}+2\nu\log{\nu\over a^2}$, see fig.~\ref{fi:loop1}. In this
phase $\xi_0>\xi_+$ i.e. (for the real choice of all necessary parameters) the cut
between $\xi_-$ and $\xi_+$ \rf{sora} is located to the left from the double point
$\xi_0$.
  \item  Growing droplet after the "impurance": this phase corresponds to
${\cal X}>1+{2\sqrt{\nu}\over\Delta}$ and $t_0=t_0^{(-)}$. The liquid of
eigenvalues fills the impurance at $z=a$ and continues
to grow smoothly with growth of $t_0$, see fig.~\ref{fi:loop2}. Now $\xi_0<\xi_-$
i.e. the cut in \rf{sora} is located to the right from the double point.
\end{itemize}

\begin{figure}[tp]
\centerline{\epsfig{file=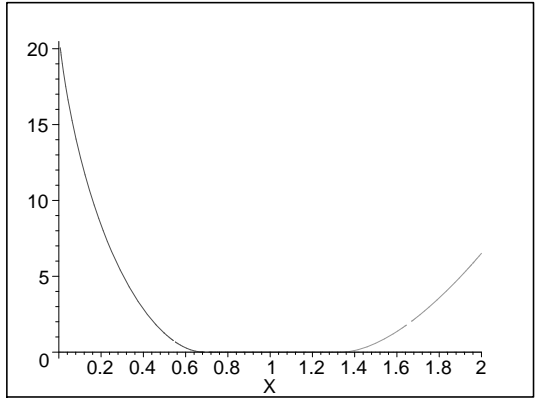,width=70mm}}
\caption{Numeric values of the integral $\Re \oint \tilde zdz$ in different phases:
$\xi_0 >\xi_+$ and $\xi_0 <\xi_-$ (for $a=5$ and $\nu=\ha$). It is seen on this picture
that they can be easily sewed for different phases, say for $\Re \oint \tilde zdz = 0$.}
\label{fi:period}
\end{figure}

Between these two phases it has to pass through the cusp. The cusp point can be
"overcome" by the solution in spirit of \cite{GP},
passing from rational curve out to nonsingular curve (\ref{jcur}) of genus $g=1$. The
degenerate genus $g=0$ solutions can be glued through the $g=1$ solution
only when they have equal values of
\be\label{repe}
\Re \oint \tilde zdz = \left\{
\begin{array}{c}
 \Re \int_{w_0^-}^{w_0^+} \tilde zdz, \ \ \ \ \ \xi_0>\xi_+ \\
 \Re \int_{w_0^+}^{w_0^-} \tilde zdz, \ \ \ \ \ \xi_0<\xi_-
\end{array}\right.
\ee
where $w_0^\pm\equiv \left.w^\pm\right|_{z=\xi_0}$ are two values of the pullback
of the point $z=\xi_0$ to the $w$-plane.
One can see at fig.~\ref{fi:period}, that these values are indeed glued at
different phases, in particular when $\Re \oint \tilde zdz = 0$.

\setcounter{equation}0
\section{Complex geometry of the AdS/CFT correspondence}

Matrix models discussed in \cite{MM1} and above
present the simplest example of the gauge/string duality \cite{Pol}.
Being the "zero-dimensional" example of the gauge theory,
they propose a dual description of string theory below two
dimensions, or the "target-space" theory of two-dimensional gravity
on string world-sheets.

In contrast to matrix model the AdS/CFT conjecture \cite{Malda} deals
with the {\em four-dimensional} \4N supersymmetric Yang-Mills theory (SYM)
with the $SU(N)$ gauge group.
Again, the main contribution in ${1\over N}$-expansion \rf{expan}, where
closed string loops are suppressed, comes
from the planar diagrams when $N\to\infty$  at fixed 't Hooft coupling
$\curlywedge=g_{YM}^2N=g_sN$ (coinciding with $t_0$ for the
matrix models, we intensively used in \cite{MM1} and previous sections),
while string coupling $g_s=g_{YM}^2$ is equivalent to the
quasiclassical parameter $\hbar$.
At $\curlywedge \gg 1$ the \4N SYM theory is believed to be
dual to string theory in $AdS_5\times S^5$ with the equal, up to a sign
(positive for the sphere $S^5$ and negative for the Lobachevsky space
$AdS_5$), radii of
curvature ${R\over\sqrt{\alpha'}}=\curlywedge^{1/4}$. Therefore any
test of the AdS/CFT conjecture
implies comparing analytic series at $\curlywedge=0$
(SYM perturbation theory, which is just a direct analog of the counting of
diagrams in matrix models as in \cite{maint}) with analytic in
$\alpha'\propto{1\over\sqrt{\curlywedge}}$
worldsheet expansion.

A possible partial way-out from this discrepancy in parameters of
expansion can be to consider the {\em classical} string solutions
with large values of integrals of motion
(usually referred as "spins" $J$) on $AdS_5\times S^5$ side
\cite{GuKlePo}, whose energies
should correspond to anomalous dimensions of
"long" operators on gauge side. In this case the classical string energy
of the form $\Delta = \sqrt{\curlywedge}{\cal E}
\left({J\over\sqrt{\curlywedge}}\right)$ may have an expansion of the form
$\Delta = J + \sum_{l=1}^\infty E_l\left({\curlywedge\over J^2}\right)^l$ over
the integer powers of 't Hooft coupling \cite{tseytlin}, which can be treated as series
at $\curlywedge=0$ even at $\curlywedge\gg 1$ provided large $\curlywedge$ is suppressed
by large value of the integrals of motion\footnote{On integrability
on the string side of duality see, e.g. \cite{Aru}.} $J$.
If it happens (this is not, of course, guaranteed) the classical
string energy can be tested by direct comparison with perturbative
series for the gauge theory.

Such expansion for the classical string energy can be, of course,
corrected by quantum corrections on string side. Up no now there is no
consistent way to quantize string theory in nontrivial background with
the Ramond-Ramond flux switched on. Therefore, such quasiclassical test
of the AdS/CFT correspondence should be treated rather carefully. We
shall see below, that indeed the quasiclassical picture allows to test
the conjecture in the first non-vanishing orders, but more detailed
analysis requires more knowledge of quantum theory on string side.

\subsection{Renormalization and the Bethe anzatz
\label{ss:bethe}}

The four-dimensional \4N SYM is conformal theory, i.e. $\beta (g_{YM})=0$,
but the anomalous dimensions $\gamma$ of the
composite operators, e.g. $\Tr \left(\Phi_{i_1}\dots\Phi_{i_L}\right)(x)$
(where $x$ is some point of the four-dimensional space-time)
are still renormalized nontrivially. Below we consider
the particular scalar operators from this set, though the proposed
approach \cite{KMMZ} can be applied in much more general situation. On string side
such operators correspond to the string motion in the compact $S^5$-part
of ten-dimensional target-space, due to standard Kaluza-Klein argument.
To simplify the situation maximally,
choose two complex $\bPhi_1=\Phi_1+i\Phi_2$ and $\bPhi_2=\Phi_3+i\Phi_4$
fields
among six real $\Phi_i$ and consider the {\em holomorphic} operators
\be
\label{holop}
{\rm Tr}\left( \bPhi_1 \bPhi_1 \bPhi_1 \bPhi_2 \bPhi_2 \bPhi_1 \bPhi_2 \bPhi_2
\bPhi_1 \bPhi_1 \bPhi_1 \bPhi_2 \ldots\right)(x)
\ee
which can be conveniently labeled by arrows as
$\left|\uparrow\uparrow\uparrow\downarrow\downarrow\uparrow
\downarrow\downarrow\uparrow\uparrow\uparrow\downarrow\ldots
\right\rangle \in \left(\mathbb{C}^2\right)^{\otimes L}$.

These operators do not have singularities at coinciding arguments of all
fields, since the kinetic terms for the holomorphic fields do always have
the $\bar\bPhi\bPhi$-structure.
The holomorphic subsector is "closed" under renormalization and
the anomalous dimensions are eigenvalues of the $2^L\times 2^L$ mixing
matrix
\be
\label{hahe}
H=\frac{\curlywedge}{16\pi^2}\sum_{l=1}^L
\left(1-{\sigma}_l\cdot{\sigma}_{l+1}\right)+O(\curlywedge^2)
\ee
which is, up to addition of a constant, the permutation operator in
$\left(\mathbb{C}^2\right)\otimes
\left(\mathbb{C}^2\right)$, whose appearance is determined by structure of the
$\Phi^4$-vertex in SYM Lagrangian,
or the Hamiltonian for Heisenberg magnetic \cite{MiZa}.

It is well-known, that the matrix \rf{hahe}
can be diagonalized using the Bethe anzatz \cite{Bethe}, (see e.g.
\cite{Faddeev} for present status of this technique and
comprehensive list of references).
Integrable structure of Heisenberg spin chain is encoded in the
transfer matrix
\be
\label{trama}
\hat{T}(u)={\rm Tr}\,\left[\left(u
+\frac{i}{2}\,{\sigma}_L\otimes{\sigma}\right)
\ldots\left(u
+\frac{i}{2}\,{\sigma}_1\otimes{\sigma}\right)\right]
\ee
which can be presented
as an operator-valued matrix in the auxiliary two-dimensional space
\be
\label{TABCD}
\hat{T}(u)
= \Tr\left(
\begin{array}{cc}
  \hat A(u) & \hat B(u) \\
  \hat C(u) & \hat D(u) \\
\end{array}\right) =
\hat A(u) + \hat D(u)
\ee
with four "generators of Yangian"
$\hat A(u)$, $\hat B(u)$, $\hat C(u)$ and $\hat D(u)$, satisfying the
quadratic RTT-algebra.
Expansion of the operator \rf{trama} in spectral parameter
at $u={i\over 2}$ generates the local charges
\be\label{tu}
\hat{T}(u)=\left(u+\frac{i}{2}\right)^L
\hat{U} \exp\left[i\sum_{n=1}^{\infty}
\frac{1}{n}\left(u-\frac{i}{2}\right)^n\hat{Q}_n\right]
\ee
where $\hat{U}=e^{i\hat{P}}$ is the shift operator (by one site along the chain)
and the Hamiltonian \rf{hahe} $\hat{Q}_1\propto H$ is one of the local charges.
Contrarily, expansion of \rf{trama} at $u=\infty$ produces the non-local
Yangian charges, whose role in the context of \4N SYM was discussed in \cite{DNW}.

Bethe ansatz diagonalizes the {\em whole} spectral-parameter dependent operator
\rf{trama}, \rf{tu}
\be
\label{tegnv}
\hat{T}(u)\left|u_1\ldots u_J\right\rangle
=T(u)\left|u_1\ldots u_J\right\rangle,
\ee
explicitly constructing the eigenvectors $\left|u_1\ldots u_J\right\rangle$ in
terms of the operators \rf{TABCD}
\be
\label{bvect}
\left|u_1\ldots u_J\right\rangle = \hat B(u_1)\ldots\hat B(u_J)
\left|\uparrow\uparrow\uparrow\uparrow\uparrow\ldots
\right\rangle
\ee
acting on the ferromagnetic vacuum
$\left|\uparrow\uparrow\uparrow\uparrow\uparrow\ldots
\right\rangle$. Any Bethe vector \rf{bvect}
depends on $J$ rapidities, and the corresponding
eigenvalues of the operator \rf{tegnv} are
\be\label{eigenv}
T(u)=\left(u+\frac{i}{2}\right)^L\prod_{k=1}^J\frac{u-u_k-i}{u-u_k}\
+ \left(u-\frac{i}{2}\right)^L
\prod_{k=1}^J\frac{u-u_k+i}{u-u_k}
\ee
Now since any eigenvalue $T(u)$ of the operator \rf{trama}
must be a polynomial in $u$, the cancelation of
poles at $u=u_j$ in the expression \rf{eigenv} gives $J$ conditions:
\be\label{BAEQ}
\left(u_j+i/2\over u_j-i/2\right)^L=
\prod_{k=1(k\ne j)}^J {u_j-u_k+i\over u_j-u_k-i},
\ee
which are the Bethe equations.

Hence, the eigenvectors of Hamiltonian \rf{hahe}
are parameterized by Bethe roots $\{ u_1,\dots, u_J\}$, ($J \leq \half L$
due to an obvious $\mathbb{Z}_2$-symmetry, in particular permuting two ferromagnetic
vacua $\left|\uparrow\uparrow\uparrow\uparrow\uparrow\ldots
\right\rangle \leftrightarrow \left|\downarrow\downarrow\downarrow\downarrow
\downarrow\ldots
\right\rangle$), satisfying \rf{BAEQ}. After taking
the logarithm, eqs.~\rf{BAEQ} acquire the form
\be\label{BAE}
L\log\left(u_j+i/2\over u_j-i/2\right)=2\pi in_j
+\sum_{k\ne j}^J \log{u_j-u_k+i\over u_j-u_k-i}
\ee
with appeared mode numbers $n_j\in\mathbb{Z}$ being some integers, never vanishing
simultaneously for a nontrivial solution. For the diagonalization of
the mixing matrix of the operators \rf{holop} in SYM theory, equations \rf{BAE}
are supplied by the "trace condition"
\be
\label{MOMu}
e^{iP}=\prod_{j=1}^J{u_j+i/2\over u_j-i/2}=1,
\ee
i.e. integrality of the total momentum of magnons: ${P\over 2\pi}\in \mathbb{Z}$.

The energy of the eigenstate of the Hamiltonian \rf{bvect} is given by sum
over magnons, i.e.
$\propto \sum_{j=1}^J \left.{dp\over du}\right|_{u=u_j}$, for
the quasimomentum
$\exp ip(u) = {u+i/2\over u-i/2}$, and it means that
the anomalous dimensions for the mixing matrix \rf{hahe} in the first order
of the SYM perturbation theory can be written as
\be
\label{gamad}
\gamma={\curlywedge\over 8\pi^2}\sum_{j=1}^J{1\over u_j^2+1/4} + O(\curlywedge^2)
\ee
and these $\gamma$'s one needs for testing the quasiclassical AdS/CFT conjecture.

\begin{figure}[tp]
\centerline{\epsfig{file=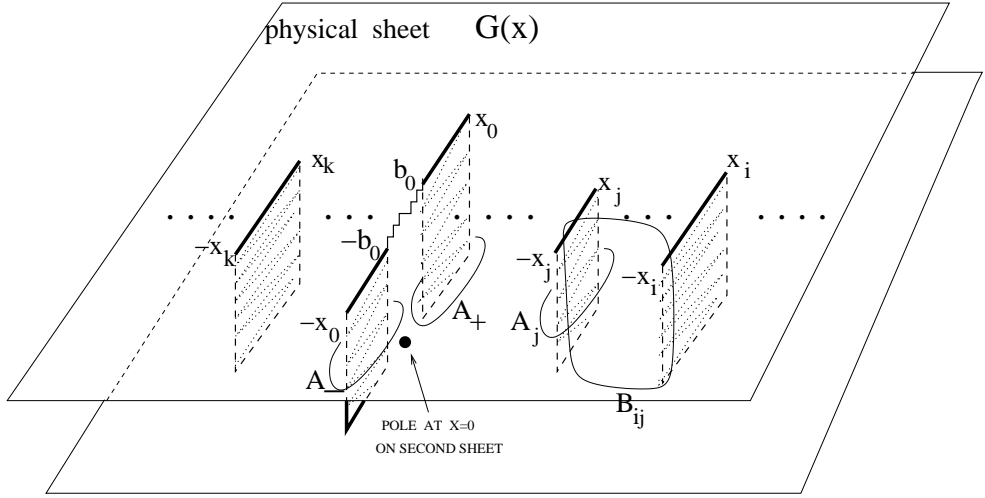,width=130mm,angle=0}}
\caption{Riemann surface $\Sigma$, which is a double-cover of the $x$-plane
cut along the Bethe strings (four slightly curved lines on each sheet in this
example), which cross the real axis at $x={1\over 2\pi n_l}$. There can be also
possible cuts along the imaginary axis, which can be however eliminated after
appropriate choice of the homology basis in $H^1(\Sigma)$, see \cite{KMMZ}.
The upper sheet is physical for the Abelian integral $G(x)$, while on the
lower sheet the resolvent has extra singularities.}
\label{fi:sigma}
\end{figure}

For comparison with dual (classical) string theory we are interested in the
long operators with $L\to\infty$, for which it is known empirically that
the Bethe roots are
typically of the order of $u_j\sim L$. Rescaling $u_j = Lx_j$, and
omitting higher in ${1\over L}$ terms, one gets from \rf{BAE} the equations of more
simple form
\be\label{BAEG}
{1\over x_j}=2\pi n_j+ {2\over L}\sum_{k\ne j}^J {1\over
x_j-x_k}
\ee
Like in the matrix model case (see detailed discussion in
\cite{MM1}), if the second "interaction term" in the r.h.s.
is absent, the solution to \rf{BAEG} is given by
$x_j= {1\over 2\pi n_j}$ for each $n_j$ (filling the extrema!), and
when one switches on the interaction,
the roots corresponding to $n_j$ (if there are several for each value)
will "concentrate" around
${1\over 2\pi n_j}$ forming the so called "Bethe strings", shown at
fig.~\ref{fi:sigma}.
In other words, if several
different $x_{j_1},\dots,x_{j_m}$ have the same mode number
$n_j=n_{j_1}=\dots=n_{j_m}$,
these $m$ Bethe roots combine into a Bethe string with the mode number
$n_{j_m}\neq 0$.

Introduce at $L\to\infty$, as in sect.~3 of \cite{MM1}, the density
\be
\label{rhoJL}
\rho(x)=\frac{1}{L}\sum_{j=1}^J\delta(x-x_j), \ \ \ \
\int_{\bf C} dx\rho(x)={J\over L}
\ee
or resolvent
\be
\label{defG}
G(x ) = {1\over L}\sum_{j=1}^J {1\over x-x_j} =
\int_{\bf C}{d\xi\,\rho(\xi)\over x-\xi}, \ \ \ \
{1\over 2\pi i}\oint_{\bf C} dxG(x)={J\over L}
\ee
and consider the {\em finite} number of different Bethe strings
$n_l\neq n_{l'}$, with $l,l'=1,\dots,K$ despite the infinitely many roots
$J$. Then in the scaling limit ($J\to\infty$ with fixed
finite $K$) the total eigenvalue support is
${\bf C}={\bf C}_1\bigcup\ldots\bigcup{\bf C}_K$, where on each component
one gets from \rf{BAEG}
\be\label{BAEC}
2\vpint_{{\bf C}} {d\xi\,\rho(\xi)\over x-\xi} =
G(x_+)+G(x_-) = {1\over x}-2\pi n_l,\ \ \ x\in
{\bf C}_l
\ee
where $G(x_\pm)$ are values of the resolvent on two different sides of the
cut along the Bethe string.

The integrality of total momentum condition, using the Bethe
equations \rf{BAEG}, acquires the form
\be
\label{MOMx}
{1\over L}\sum_{j=1}^J {1\over x_j} =
2\pi \sum_{l=1}^K n_l \int_{{\bf C}_l}\rho(x)dx =
2\pi m, \ \ \ \ m\in \mathbb{Z}
\ee
or
\be
\label{MOMrho}
{1\over 2\pi i}\oint_{{\bf C}}{G(x)dx\over x}
= 2\pi m, \ \ \ \ \ n_l, m\in \mathbb{Z}
\ee
Different $n_l\neq n_{l'}$ on different parts of support ${\bf C}_l\bigcap{\bf
C}_{l'}=\emptyset$ mean that, in contrast to the matrix model case,
discussed in detail in sect.~3 of \cite{MM1},
$G(x)=\int^x dG$ is not already a single-valued function, but an
Abelian integral on some hyperelliptic curve $\Sigma$, introduced by hands, according
to the number of nontrivial Bethe strings
\be
\label{sigmaxxx}
 y^2 = R_{2K}(x) = x^{2K} +
r_1x^{2K-1} + \dots + r_{2K} =
\prod_{j=1}^{2K} (x-{\rm x}_j)
\ee
It means, that resolvent $G(x)$ in contrast to the case of one-matrix model
does not satisfy here a sensible algebraic
equation, but can be still determined from its geometric properties.

Equations \rf{defG}, \rf{BAEC} and \rf{MOMrho} can be solved after
reformulating them as a set of properties of the meromorphic
differential $dG$ of the resolvent \rf{defG}:
\begin{itemize}
  \item The differential $dG$ is the second-kind Abelian differential with the only
second-order pole at the point $P_0$, ($x(P_0)=0$ on unphysical
sheet of the Riemann surface $\Sigma$, see fig.~\ref{fi:sigma}), in particular
it means that $\oint_{A_i}dG = 0$ for all $A$-cycles, surrounding the
Bethe strings. The absence of poles on physical sheets is determined by properties
of the resolvent, but on unphysical sheets then any differential with vanishing
$A$-periods must have singularities;
  \item In addition to vanishing $A$-periods,
  the differential $dG$ has integral $B$-periods
\be
\label{Bint}
\oint_{B_i}dG = 2\pi (n_i-n_K)
\ee
More exactly one can write \cite{KMMZ}
\be
\label{B'int}
\int_{B'_j}dG = 2\pi n_j, \ \ \ \ j=1,\dots,K+1
\ee
where $B'_j$ is the contour from $\infty_+$ on the upper sheet to
$\infty_-$ on the lower sheet, passing through the $j$-th cut, so that
$B_j=B'_j-B'_K$, for $j=1,\dots,K$, see fig.~\ref{fig:openC}.
\begin{figure}[tp]
\centerline{\epsfig{file= 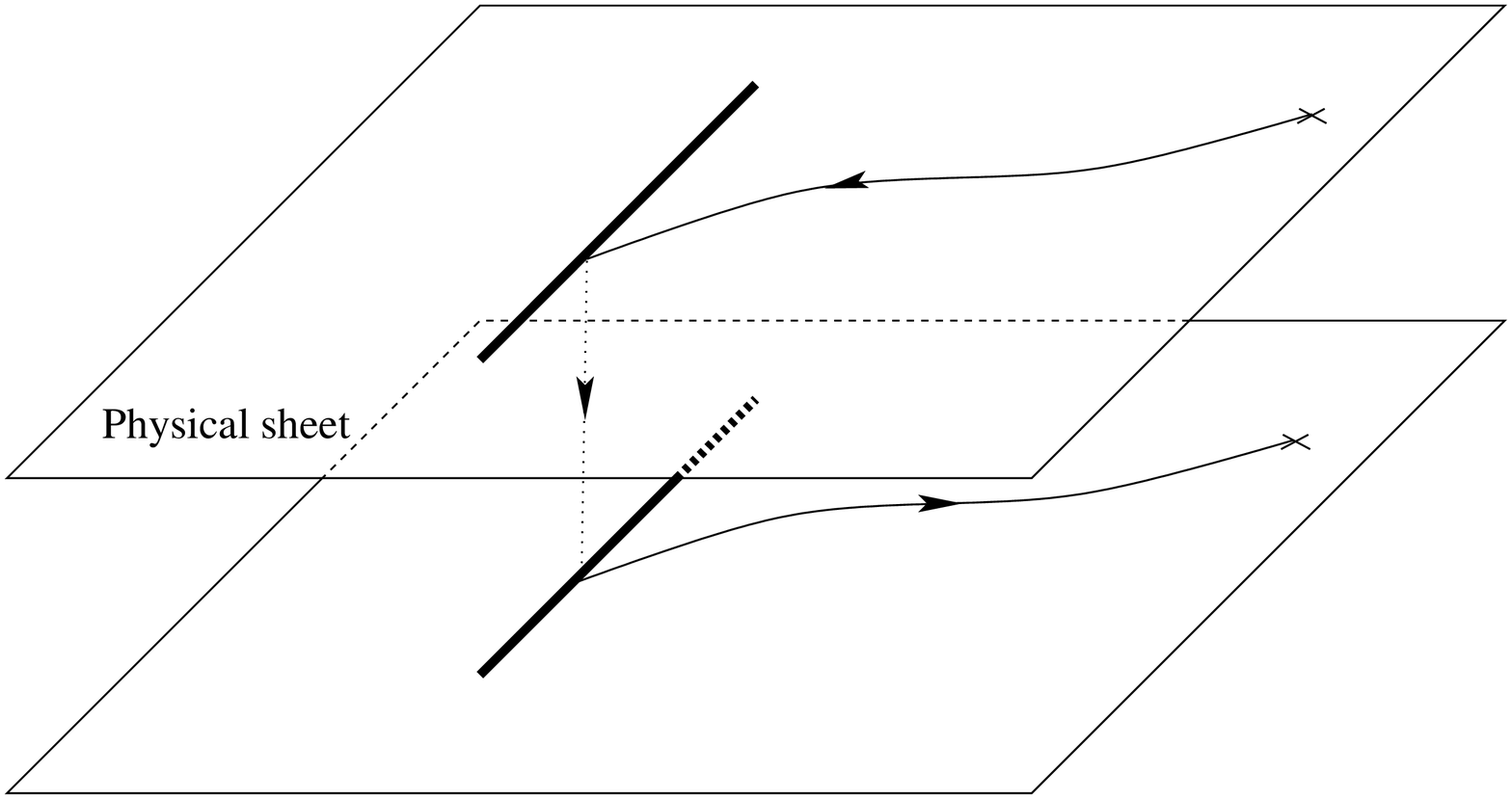,width=90mm}}
\caption{The integration contour $B'$ in formula \rf{B'int}: the marked points
on both sheets correspond to infinities $\infty_+$ and $\infty_-$ where
$x=\infty$.}
\label{fig:openC}
\end{figure}

  \item $dG$ has the following behavior at infinity
\be
\label{inf}
dG\ \stackreb{x\to\infty}{=}\ {J\over
L}{dx\over x^2}+\dots
\ee
and the Abelian integral $G(x)$ itself is fixed by \rf{MOMrho}
\be
\label{zero}
G(x) = 2\pi m + \int_0^x dG, \ \ \  {\rm or}\ \  G(0)=2\pi m
\ee
\end{itemize}
The general solution \cite{KMMZ} for the differential $dG$ on hyperelliptic curve
\rf{sigmaxxx}, satisfying the above requirements
\rf{Bint}, \rf{B'int}, \rf{inf} and \rf{zero} is read from a standard formula for the
second-kind Abelian differential on hyperelliptic curve
\be
\label{gsol}
dG =
-{dx\over 2x^2}\left(1-{\sqrt{r_{2K}}\over y}\right)
+ {r_{2K-1}\over 4\sqrt{r_{2K}}}{dx\over xy}
+ \sum_{k=1}^{K-1}a_k {x^{k-1}dx\over y}
\ee
together with the extra conditions, ensuring, in particular, the
single-valuedness of the resolvent on "upper" physical sheet
\be
\label{Aint}
\oint_{A_i}dG = 0,\ \ \ \   i=1,2,\ldots,K-1
\ee
which is a system of linear equations,
to be easily solved for the coefficients $\{ a_k \}$.
The rest of parameters is "eaten by" fractions of roots on particular
pieces of support
\be
\label{frac}
S_j  = \int_{{\bf C}_j} \rho(x)dx
= -{1\over 2\pi i}\oint_{A_j}x dG, \\ j=1,\dots,g=K-1
\ee
the total amount of Bethe roots \rf{rhoJL},
and the total momentum \rf{MOMrho}.

The energy or one-loop anomalous dimension for generic finite-gap
solution \cite{KMMZ} can be read from \rf{gamad}, \rf{gsol}
\be
\label{gamma}
\gamma = {\curlywedge\over 8\pi^2 L}\oint_{\bf C} {dx\over 2\pi i\, x^2}\, G(x)=
{ \curlywedge\over 8\pi^2L  }\left({r_{2K-2}\over 4 r_{2K}}
-\frac{r^2_{2K-1}}{16r^2_{2K}}
-{a_1\over \sqrt{r_{2K}}}\right)
\ee
The anomalous dimensions defined by \rf{gamma} are functions
of the coefficients of the embedding equation \rf{sigmaxxx} and
$a_1$ which again
is expressed through these coefficients by means of \rf{Aint}. The
moduli of the curve \rf{sigmaxxx} are themselves
(implicitly) expressed through the mode numbers $n_j$ and
root fractions
$S_j$ via \rf{Bint} or \rf{B'int} and \rf{frac} (together
with the total momentum
\rf{MOMrho} and the total number of Bethe roots \rf{defG}).

It is interesting to point our, that formula \rf{gsol} is a particular case
of the two-point resolvent for the planar matrix model $W_0(x,\xi)$, or
so called Bergman kernel $\omega(P,P')=d_Pd_P'\log E(P,P')$ (see also
formula \rf{bidif} from Appendix~\ref{app:Fay}), which on a
(tensor square of) the  hyperelliptic curve \rf{sigmaxxx} reads
\be
\label{bekehy}
W_0(x,\xi)dx d\xi=
\\
-{dx d\xi\over 2(x-\xi)^2}+
\frac{y(x )}{2y(\xi)}\left(\frac{1}{(\xi-x )^2}+
\frac12\sum_{\alpha=1}^{2n}\left[\frac{1}{(\xi-x )(x -{\sf x}_\alpha)}
-\sum_{i=1}^{n-1}{ H}_i(\xi)
\oint_{A_i}\frac{dx}{(x-{\sf x}_\alpha)^2{y (x)}}\right]\right)dx d\xi =
\\
=-{dx d\xi\over 2(x -\xi)^2}\left(1-{R(x)\over
y(x )y(\xi)}\right) - {dx d\xi\over 2(x -\xi)}
{R'(x)\over y(x )y(\xi)}
- {1\over 2}y(x)dx\sum_{\alpha=1}^{2n}\sum_{i=1}^{n-1}d\omega_i(\xi)
\oint_{A_i}\frac{dx}{(x-{\sf x}_\alpha)^2{y (x)}}
\ee
The detailed discussion of this issue can be found in \cite{ChMMV2}.

\subsection{Geometry of integrable classical strings}

Formulas \rf{gsol} - \rf{frac} of the previous paragraph show, that
the general solution for anomalous dimension of long operators \rf{gamma} is
expressed through the
integrals of motion on some {\em classical} configurations of the Heisenberg
magnet\footnote{Such correspondence with classical solutions
was first noticed in \cite{RESHSM}
for the non-linear Schr\"odinger equation.}.
In the dual string picture one has the classical trajectories of
string,
moving in (some subspace of) $AdS_5\times S^5$ and the finite gap
solutions to string sigma-model in the Lobachevsky-like spaces were first
constructed in \cite{Krisig}. Being slightly modified, it can be easily
applied to the case of compact $S^{2D-1}$ sigma-models. In this
section, following \cite{KMMZ,MKMMZ} we are going to show how these classical
solutions can be compared with
the quasiclassical solutions on the gauge side.

In particular subsector of only two holomorphic fields one gets the
$S^3\subset S^5$ sigma-model (in the $AdS_5$-sector the only nontrivial
string co-ordinate on the solution is "time"
$X_0={\Delta\over\sqrt{\curlywedge}}\ \tau$) with the action
\be
\label{sigmamo}
S ={\sqrt{\curlywedge}\over 4\pi }\int d\sigma d\tau
\left[ \left(\d_aX_i\right)^2 -\left(\d_a X_0\right)^2\right],
\ee
($\sum X_i^2=1$). Provided by identifications (since $S^3$
is the group-manifold of the Lie group $SU(2)$)
\be
\label{su2}
g = \left(
\begin{array}{cc}
  X_1+iX_2 & X_3+iX_4 \\
 -X_3+iX_4 & X_1-iX_2
\end{array}
\right) \equiv
\left(
\begin{array}{cc}
  Z_1 & Z_2 \\
  -{\bar Z}_2 & {\bar Z}_1
\end{array}\right) \in SU(2),
\\
J=g^{-1}dg = \left(
\begin{array}{cc}
  {\bar Z}_1dZ_1 + Z_2d{\bar Z}_2 &  {\bar Z}_1dZ_2 - Z_2d{\bar Z}_1 \\
   {\bar Z}_2dZ_1 - Z_1d{\bar Z}_2 &  {\bar Z}_2dZ_2 + Z_1d{\bar Z}_1
\end{array}\right) \in su(2).
\ee
it is equivalent \cite{FR}
to the $SU(2)$ principal chiral field  with the Lax pair,
\be
J_\pm(x) =
{\Delta\over\sqrt{\curlywedge}}\frac{i{\bf S}_\pm\cdot\mathbf{\sigma}}{1\mp X}
\\
\d_+J_--\d_-J_++[J_+,J_-]=0
\\
\d_+J_-+\d_-J_+=0
\label{zerocur}
\ee
which has two simple poles at values of string spectral
parameter $X= X(P_\pm)=\pm {\sqrt{\curlywedge}\over 4\pi\Delta}$.
In different
words, such sigma-model is equivalent
to a system of {\it two} interacting relativistic spins
${\bf S}_+$ and ${\bf S}_{-}$:
\be
\label{emcf}
\d_+{\bf S}_{-}+{2\Delta\over\sqrt{\curlywedge}}\ {\bf S}_{-}\times{\bf S}_{+}=0,
\\
\d_-{\bf S}_{+}-{2\Delta\over\sqrt{\curlywedge}}\ {\bf S}_{-}\times{\bf S}_{+}=0.
\ee
which
in the "non-relativistic limit"
\be\label{suv}
{\bf S}_\pm = {\bf S}\pm\frac{\sqrt{\curlywedge}}{4\Delta}\,{\bf S}\times{\bf S}_\sigma
-\frac{\curlywedge}{32\Delta^2}\left({\bf S}_\sigma^2\right)\,\,{\bf S}
+\ldots
\ee
degenerates into the Heisenberg magnet, similar to the procedure,
studied in the papers \cite{Kruczenski}.

\begin{figure}[tp]
\centerline{\epsfig{file=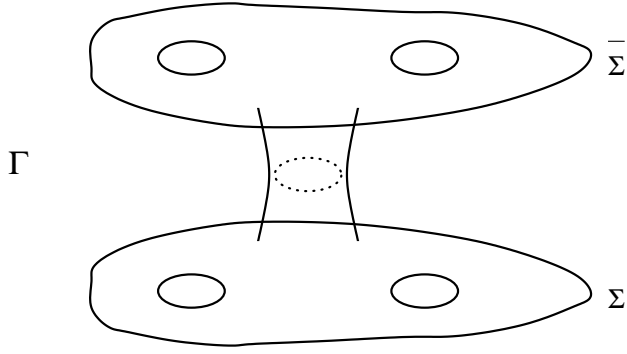,width=75mm,angle=-90}}
\caption{Riemann surface $\Gamma$, which is a double cover of $\Sigma$
with a single cut. It has an involution exchanging $\Sigma \leftrightarrow \bar\Sigma$.}
\label{fi:gamma}
\end{figure}

However, this method can be used only for the group-manifolds. Nevertheless,
in general situation the string sigma-model solution for the complex
co-ordinates
$Z_I(\tau,\sigma)$ and $\bar Z_I(\tau,\sigma)$ on $S^{2D-1}$, (constraint
by $\sum_I |Z_I|^2=1$)
\be
\label{zkri}
Z_I(\sigma_\pm) = r_I\Upsilon(q_I,\sigma_\pm)
\ \ \ \ \
{\bar Z}_I(\sigma_\pm) = r_I\Upsilon({\bar q}_I,\sigma_\pm)
= r_I\overline{\Upsilon(q_I,\sigma_\pm)},
\ \ \ \ \
I=1,\dots,D
\ee
($\sigma_\pm = \half(\sigma\pm \tau)$) can be found \cite{Krisig}
in terms of the Baker-Akhiezer (BA) functions
\be
\label{BAsigma}
\Upsilon(P,\sigma_\pm)\
\stackreb{P\to P_\pm}{=}\ e^{k_\pm \sigma_\pm}
\left(1+\sum_{j=1}^\infty {\xi_j(\sigma_\pm)\over k_\pm^j}\right)
\propto e^{\Omega_+(P)\sigma_+
+ \Omega_-(P)\sigma_-}
\theta \left(\bomega(P)+ {\bf V}_+\sigma_+
+ {\bf V}_-\sigma_-\right)
\ee
where, $\bomega(P)$ is the Abel map \rf{E1s} and $\theta \left(\bomega\right)$ is
the Riemann theta function \rf{thetadef} (see Appendix~\ref{app:Fay} for details),
defined on double cover $\Gamma$ (branched at $P_+$ and $P_-$)
of a Riemann surface $\Sigma$ (see fig.~\ref{fi:gamma}).
For only two complex co-ordinates $Z_I$ (like in the $S^3$
case) the curve $\Sigma$ is hyperelliptic and directly related with the
curve \rf{sigmaxxx} of the Heisenberg chain.

\begin{figure}[tp]
\centerline{\epsfig{file=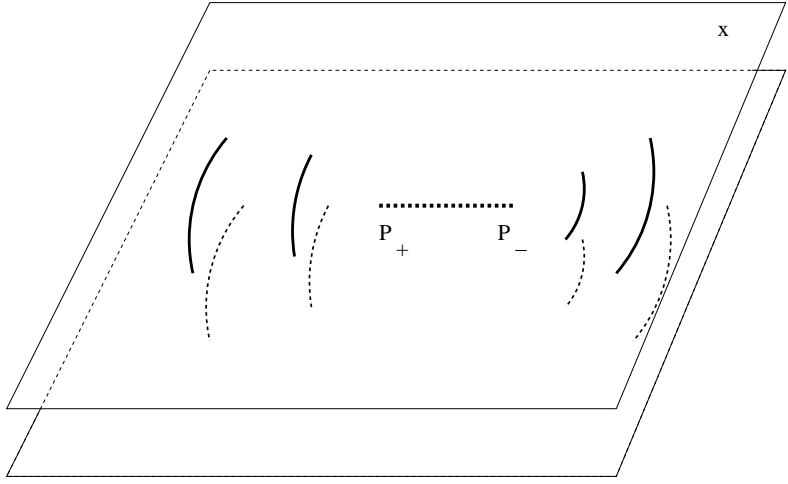,width=75mm,angle=-90}}
\caption{Riemann surface $\Sigma$ for classical string theory. Compare to
fig.~\ref{fi:sigma}, the pole on unphysical sheet is replaced by a cut between $P_+$
and $P_-$ of the length ${\sqrt{\curlywedge}\over 2\pi\Delta}$. Two copies of this
Riemann surface, glued along this extra cut give rise to the full string curve $\Gamma$
from fig.~\ref{fi:gamma}.}
\label{fi:sigmastr}
\end{figure}

More strictly, the string hyperelliptic curve $\Sigma$ turns into the spin chain
hyperelliptic curve \rf{sigmaxxx}, depicted at fig.~\ref{fi:sigma} only in the limit
$\curlywedge/\Delta^2 \to 0$, consistently with \rf{suv}. Schematically it
can be drawn as on fig.~\ref{fi:sigmastr}, which is quite similar to the spin chain
curve from fig.~\ref{fi:sigma} except for the pole on unphysical sheet, which is now
replaced by an extra cut of the length ${\sqrt{\curlywedge}\over 2\pi\Delta}$.

The BA function \rf{BAsigma} (and hence the solution to sigma-model)
is constructed in terms of {\em two}
second-kind Abelian differentials $d\Omega_\pm$ on Riemann surface $\Gamma$
\be
d\Omega_\pm\ \stackreb{P\to P_\pm}{=}\
\pm dk_\pm\left(1+O(k_\pm^{-2})\right),
\ \ \ \ \ \oint_{\bf A}d\Omega_\pm = 0
\ee
with the only second-order pole at
$P_\pm$ respectively;
${\bf V}_\pm = \oint_{\bf B} d\Omega_\pm$ are the vectors of
their $B$-periods.

The proof of the fact that formulas \rf{zkri} are solutions to the sigma-model,
satisfying classical Virasoro constraints, is based on existence of
the third-kind Abelian differential $d\Omega$ on $\Sigma$
with the simple poles at $P_\pm$ and zeroes in the poles
of the BA functions $\Upsilon$ and conjugated $\bar\Upsilon$. Then
one may define the string resolvent or quasimomentum by the following
formula
\be
\label{gstr}
d\CG = \ha\left(d\Omega_+ - d\Omega_-\right)
\\
d\CG\ \stackreb{P\to P_\pm}{=}\
\ha dk_\pm\left(1+O(k_\pm^{-2})\right),
\ \ \ \ \ \oint_{\bf A}d\CG = 0
\ee
For periodic in $\sigma$ solution, as follows from \rf{BAsigma},
the $B$-periods of the resolvent
${1\over 2\pi}\oint_{\bf B} d\CG \in\mathbb{Z}$ are integer-valued,
and for the periodic solutions one can write
\be
\label{domega}
d\Omega = {\bar\Upsilon\Upsilon \over
\langle\bar\Upsilon\Upsilon\rangle}d\CG
\ee
where brackets mean the average over the period in $\sigma$-variable.

The BA function \rf{BAsigma} satisfies the second-order
differential equation
\be
\label{wave}
(\d_+\d_- + U)\Upsilon (P,\sigma_\pm)=0, \ \ \ \ P\in\Gamma
\ee
where $U \propto
\sum_i\left(\d_+ Z_i\d_-\bar Z_i + \d_- Z_i\d_+\bar Z_i\right)$.
This fact and the Virasoro constraints
$\curlywedge\sum_I\left|\d_\pm Z_I\right|^2 = \Delta$ are guaranteed
by the properties of the differential \rf{domega} and existence
of the function $E$ on $\Sigma$ with $D$ simple poles $\{ q_I\}$ and the
following behavior at the vicinities of the points
$P_\pm$: $E=E_\pm \pm {4\pi\Delta\over\sqrt{\curlywedge}}
{1\over k_\pm^2} + \dots$, (cf. with \cite{Krisig}).

The normalization factors in the expressions for the sigma-model
co-ordinates (\ref{zkri}) are determined by the formulas
\be
\label{ris}
r_I^2 = {\res_{q_I} Ed\Omega\over E_--E_+}, \ \ \ \ I=1,\dots,D
\ee
where $E_\pm = E(P_\pm)$ and
normalizations (\ref{ris}) satisfy $\sum_{I=1}^D r_I^2=1$ due to
vanishing of the total sum over the residues $\sum \res
\left(Ed\Omega\right) = 0$.

Rescaling
$\d_\pm\to\frac{\sqrt{\curlywedge}}{4\Delta}\,\d_\tau\pm \d_\sigma$,
$\Upsilon\to e^{{2\Delta\over\sqrt{\curlywedge}}\tau}\Psi$,
$\bar\Upsilon\to e^{-{2\Delta\over\sqrt{\curlywedge}}\tau}\bar\Psi$,
$U\to U-{4\Delta^2\over\curlywedge}$,
in the limit $\curlywedge/\Delta^2 \ll 1$, one gets from \rf{wave} the
non-stationary Schr\"odinger equation
\be
\label{KP}
\left(\d_\tau-\d_\sigma^2 + U\right)\Psi = 0
\\
\left(-\d_\tau-\d_\sigma^2 + U\right)\bar\Psi = 0
\ee
where now both BA functions $\Psi$ and $\bar\Psi$ can be defined on Riemann
surface $\Sigma$ (see fig.~\ref{fi:sigma}), with the ends of the extra cut
$P_\pm$ are shrinked to a single point $P_0$, with the expansion at the vicinity
of this point (with new local parameter $k(P_0)=\infty$)
\be
\label{PsiKP}
\Psi \stackreb{P\to P_0}{=} e^{k\sigma+k^2\tau}\left(1+{\psi_1\over k}+
{\psi_2\over k^2}+
\dots\right)
\\
\bar\Psi \stackreb{P\to P_0}{=} e^{-k\sigma-k^2\tau}\left(1+
{\bar\psi_1\over k}+
{\bar\psi_2\over k^2}+\dots\right)
\ee
Substituting expansions \rf{PsiKP} into \rf{KP} one gets
\be
\label{relKP}
U = 2{\d\psi_1\over\d\sigma}=-2{\d\bar\psi_1\over\d\sigma}
\\
{\d\psi_1\over\d\tau}-2{\d\psi_2\over\d\sigma}-{\d^2\psi_1\over\d\sigma^2}+
U\psi_1=0
\\
-{\d\bar\psi_1\over\d\tau}+2{\d\bar\psi_2\over\d\sigma}-
{\d^2\bar\psi_1\over\d\sigma^2}+
U\bar\psi_1=0
\ee
For the conjugated functions one has $\psi_1+\bar\psi_1=0$ and then
${\d\over\d\sigma}\left(\psi_2+\bar\psi_2+\psi_1\bar\psi_1\right)
=-{\d U\over\d\sigma}$. Therefore, one gets an expansion
\be
\label{pp}
\bar\Psi\Psi \stackreb{P\to P_0}{=} 1 + {U\over k^2} + \dots
\ee
The differential \rf{domega} in this limit turns into
$d\Omega={\bar\Psi\Psi\over \langle\bar\Psi\Psi\rangle} dG$ and function
$E$ acquires a simple pole at $P_0$, i.e. $E \stackreb{P\to P_0}{=}
k+\dots$, if written in terms of new local parameter $k$.
From vanishing of the sum of the residues of differential $Ed\Omega$
one gets now
\be
U = \res_{P_0} Ed\Omega =
-\sum_{q_I} E{\bar\Psi\Psi\over \langle\bar\Psi\Psi\rangle} dG \propto
\sum_I \bar\Psi(q_I)\Psi(q_I)
\ee
It means, that $\Psi_I(\tau,\sigma)\propto\Psi (q_I)$ satisfy some vector
non-linear Schr\"odinger equation \cite{Cherednik}
\be
\label{vschro}
\left(\d_\tau-\d_\sigma^2 + \sum_J|\Psi_J|^2\right)\Psi_I = 0
\ee
In the case of $D=2$ the curve $\Sigma$ is hyperelliptic and one can take
the function $E$ with the only two poles, see below.
Then \rf{vschro} turns into the
ordinary non-linear Schr\"odinger equation, which can be transformed
to the Heisenberg magnetic chain \cite{ZT,FT}:
\be
\label{nlshe}
|\Psi|^2 \propto {\bf S}_\sigma^2
\\
\bar\Psi\d\Psi - \Psi\d\bar\Psi \propto \left( {\bf S}_\sigma\cdot
{\bf S}\times{\bf S}_{\sigma\sigma}\right)
\ee
and so on, which is a gauge transformation for the Lax operators.

An equivalent way to describe the classical string geometry was proposed in
\cite{KMMZ} and was based on reformulating of geometric data of the
principal chiral field \rf{zerocur} in terms of some Riemann-Hilbert problem.
The spectral problem on string side (a direct analog of the formulas
\rf{defG}, \rf{BAEC}, \rf{MOMrho} and \rf{gamma})
can be formulated in the following way.
Let $X$ and $\CG(X)$ be string spectral parameter and resolvent,
equal to the quasimomentum of the classical solution
(maybe up to an exact one-form). The spectral Riemann-Hilbert
problem on string side
can be written as \cite{KMMZ}
\be
\label{streq}
{1\over 2\pi i}\oint_{\bf C} \CG(X)dX= {J\over \Delta} + {\Delta-L\over 2\Delta}
\\
{1\over 2\pi i}\oint_{\bf C} {dX \CG(X)\over X} = 2\pi m
\\
\oint_{\bf C} {2tdX\over X^2}{ \CG(X)\over 2\pi i} =  \Delta - L
\ee
and
\be
\label{strbete}
\CG(X_+)+\CG(X_-)-2\pi n_l = {X\over X^2-t}
\ee
where we introduced the notation $t={\curlywedge\over 16\pi^2\Delta^2}$.

Comparison of the string \rf{streq}, \rf{strbete} and gauge \rf{BAEC}, \rf{MOMrho}
Riemann-Hilbert problems implies that
$x=X+{t\over X}$ together with $G(x)={\cal G}(X)$ is an {\em exact} change of
variables \cite{MKMMZ} for the quasiclassical AdS/CFT correspondence.
This in particular automatically implies the
involution $X\leftrightarrow{1\over X}$ on the string theory curves $\Gamma$,
see fig.~\ref{fi:gamma}.
Indeed, in terms of $k_\pm$ one has
\be
X \stackreb{P\to P_\pm}{=}
\pm\sqrt{t} + {1\over k_\pm}+\dots
\ee
then
\be
x =X+{t\over X}\ \stackreb{P\to P_\pm}{=} \pm
2\sqrt{t} \pm {1\over\sqrt{t}}{1\over k_\pm^2} + \dots
\ee
i.e. the function
$E = {1\over x}$ (up to an overall constant) satisfies all desired
properties for the modified construction from \cite{Krisig} we presented above,
e.g. when the cut between $P_+$ and $P_-$ on fig.~\ref{fi:sigmastr}
shrinks to a point $P_0$ on fig.~\ref{fi:sigma}, with $x(P_0)=0$,
the function $E$ acquires a simple pole at this point.

Proceeding further one gets
\be
dx=dX\left(1-{t\over X^2}\right) = {dX\over X}\left(X-{t\over X}\right)
= {dX\over X}\sqrt{x^2-4t}
\ee
and combination of the first and
third lines in \rf{streq} gives
\be
{1\over 2\pi i}\oint \CG(X)dX= {J\over \Delta} + {\Delta-L\over 2\Delta} =
{J\over \Delta} + t\oint {\CG(X)dX\over 2\pi iX^2}
\ee
or
\be
\label{nogabe}
{1\over 2\pi i}\oint dx G(x)={J\over \Delta}
\ee
The second line of \rf{streq} is then
\be
\label{momgabe}
{1\over 2\pi i}\oint {dx G(x)\over \sqrt{x^2-4t}} = 2\pi m
\ee
where the integral is taken around the cut between the points
$-2\sqrt{t}$ and $2\sqrt{t}$ in the $x$-plane,
and the third line of \rf{streq} gives
\be
\label{dgabe}
\oint {dx G(x)\over 2\pi i}
\left({x\over\sqrt{x^2-4t}}-1\right) = \Delta - L
\ee
The "string Bethe" equation on the cuts \rf{strbete}
turns now into
\be
\label{strgabe}
G(x_+)+G(x_-)-2\pi n_l =
{1\over\sqrt{x^2-4t}}
\ee
We now see from \rf{nogabe}, \rf{momgabe}, \rf{dgabe} and \rf{strgabe} that
the classical string theory spectral problem indeed is identical to the
quasiclassical Bethe equations on gauge side upon replacements
\be
\label{repl}
{1\over x}\ \rightarrow {1\over\sqrt{x^2-4t}} = {1\over x} + {2t\over x^3}
+\dots
\\
L \rightarrow \Delta
\\
\gamma \rightarrow \Delta - L
\ee
In other words, this leads to a nonlinear relation
\be
\label{strdg}
\Delta - L = \Gamma (\curlywedge,\Delta)
\ee
where $\Gamma (\curlywedge,L) = \gamma + O(\curlywedge^2)$ should be compared with
the first perturbative contributions to the
anomalous dimension of the supersymmetric gauge theory.

A simplest non-trivial example
of such relation is the solitonic limit of small number of Bethe roots,
leading to
the "modified" BMN formula \cite{KMMZ}
\be
\label{BMN}
\Delta-L=\sum_k N_k\left(
\sqrt{1+\frac{\curlywedge n^2_k}{\Delta^2}}-1\right)
\ee
for $J=\sum_k N_k$ expressed as a total amount of "positive" $n_k>0$ and
"negative" $n_k<0$ massive oscillators \cite{Metsaev}. Formulas \rf{strdg} and
\rf{BMN} show, that the solution for $\Delta$ of classical string theory is
given in terms of the highly non-linear formulas, and the oscillator
language of \cite{Metsaev,BMN} is rather an effective tool for description
of certain quasiclassical modes of an integrable string model in pp-wave
geometry, than an exact world-sheet quantization of the theory.

\subsection{Beyond the SU(2) subsector: overparameterized curves}

It turns out to be very hard to write down an algebraic equation for the curve
beyond the $SU(2)$ subsector. Up to now, the only achievement in this
direction was related with the highly "overparameterized" curves, i.e. determined
by equations, giving rise to much higher genera and much larger set of parameters,
which is necessary for solving the problem. These curves should be further constrained
in order to fit with the moduli space of necessary dimension for the
desired solution. The situation here is a
bit similar to the two-matrix model, see discussion of the complex curve of two-matrix
model in \cite{MM1}.

The simplest example of such overdetermined situation is given by the so called Baxter
curves, which are determined by the Baxter equation
\be\label{Baxter}
e^{ip}+e^{-ip} = \Tr\ \Omega(x)
\ee
for the quasimomentum $p=G-{1\over 2x}$, which define the finite genus
curve only for the chains with
finite number of sites $L$, this was used for similar purposes in \cite{Korch}.
Otherwise, if $L\to\infty$, \rf{Baxter} corresponds
literally to a complex curves of an infinite genus, and cannot be used
effectively for any of
practical purposes. Differently, the algebraic function $p$ has essential
singularities on the "minimal" curves, like \rf{sigmaxxx}, and it is not
generally algebraic, unless the corresponding
equation gives rise to an overparameterized curve of an infinite genus, where the
essential singularity if "dissolved".

Moreover, instead of \rf{sigmaxxx} one can write down an equation satisfied by the
meromorphic functions $p'\equiv x^2{dp\over dx}$, or ${dG\over dx}$ for the
resolvent \rf{gsol}. Clearly, it is has the form
\be
\label{p'curve}
P_2(x)(p')^2+P_0(x)=0
\ee
where $P_2(x)=R_{2K}(x)$ and $P_0(x)$ are polynomials of degree $2K$. Note, however,
that equation \rf{p'curve} corresponds to a curve of genus ${\hat g}=2K-1$, instead
of $g=K-1$ for the minimal curve \rf{sigmaxxx}, i.e. it has $K$ extra "false"
handles with corresponding extra parameters.
We shall refer to the curves of the type \rf{p'curve} as to "overparameterized" curves,
being "in between" the minimal and the Baxter ones.

\begin{figure}[tp]
\centerline{\epsfig{file=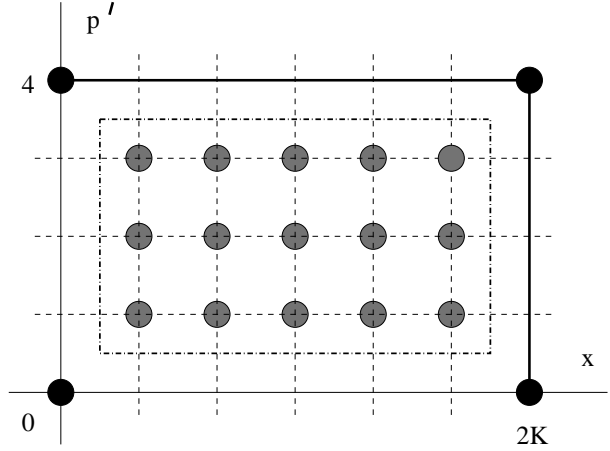,width=80mm}}
\caption{Newton polygon for the non-degenerate curve, defined by the equation
\rf{bks}. The smooth genus ${\hat g}$ given by the number of integer points
inside the polygon is ${\hat g}=3(2K-1)=6K-3$.}
\label{fi:np4}
\end{figure}

Beyond the $SU(2)$ case the overdetermined curves of the type \rf{p'curve} were
proposed in \cite{BKS} in the form of quartic equation\footnote{A similar analysis
for the most general case of octic equation can be found in \cite{SN2004ik}.}
\be
\label{bks}
P_4(x)(p')^4+P_2(x)(p')^2+P_1(x)p'+P_0(x)=0
\ee
with all polynomials $P_i(x)$, $i=1,\dots,4$ of degree $2K$. It is easy to see from
the corresponding Newton polygon, see fig.~\ref{fi:np4},
that the genus of \rf{bks} is ${\hat g}=6K-3$.

If one additionally requires, however, the
only $2K$ branch points for the curve, defined by
quartic equation \rf{bks}, the Riemann-Hurwitz formula $2-2g=l(2-2g_0)-\# B.P.$ gives
\be
\label{RHbks}
2-2g=4(2-0)-2K = 8-2K, \ \ \ \ {\rm or}
\\
g=K-3
\ee
i.e. one has to constraint ${\hat g}-g=5K$ parameters. In \cite{BKS} these constrains
are chosen in the form, suggested by comparison of \rf{p'curve} with
the minimal curve \rf{sigmaxxx}.

\setcounter{equation}0
\section{Conclusion}

In this paper, as in its first part \cite{MM1}, we have concentrated
mostly on the geometric properties of the
quasiclassical matrix models, closely related with their integrability.
The corresponding geometric data are encoded in one-dimensional
complex manifold - a complex curve or Riemann surface,
endowed with a meromorphic generating one-form.
The free energy of matrix model is given by a prepotential, or quasiclassical
tau-function, completely determined
by these geometric data. The geometric definition of the quasiclassical tau-function
satisfies the simplest possible integrability relations -
the symmetricity of its second derivatives, guaranteed by the Riemann bilinear
relations for the meromorphic differentials on this curve. The derivatives
of prepotential can be express in terms of residues, periods and
other invariant analytic structures, and as a consequence, the quasiclassical free
energy of matrix models satisfies certain non-linear integrable differential
equations, which were studied in the paper.

In this respect an immediate question arises, concerning the rest of the expansion
\rf{expan}: is there any natural geometric picture for the correlators of
matrix models beyond the
planar limit? It does not have any clear answer at the moment, except maybe for the
string one-loop correction $F_1$, is given by certain determinant formulas,
as \rf{F1}, which in certain sense
describe the fluctuations of the quasiclassical geometry,
see also \cite{F1,ChMMV2}. As for the whole sum \rf{expan}, it is rather described by a
{\em quantization} of the presented above geometric picture, and some step towards
this direction were discussed, say, in \cite{MSShish}. From the point of view of
integrable systems quantization of geometry of one-matrix model corresponds to the
so called stationary problem in the Toda chain hierarchy \cite{statoda}, while
for the two-matrix
model case one has to consider the same problem in the context of the
full hierarchy for the two-dimensional Toda lattice.

Even more questions arise concerning the quantum generalization of quasiclassical
picture of the AdS/CFT correspondence (see discussion of some of these
issues in \cite{AC1l}). While one can certainly move in this direction similarly to
studying matrix models within the $1/N$-expansion
(here the same role in some sense is played by the finite-size
corrections to the quasiclassical solutions of Bethe equations), the understanding
of the full quantum
picture necessarily requires at least some progress in formulation of quantum string
theory in non-trivial backgrounds. We believe that such progress is possible, but
this is certainly beyond the scope of particular questions, discussed in this paper.

\bigskip\bigskip\noindent
I am grateful to L.Chekhov, V.Kazakov, I.Krichever, A.Losev, J.Minahan,
A.Mironov, A.Morozov, N.Nekrasov, D.Vasiliev, A.Zabrodin, Al.Zamolodchikov and
K.Zarembo
for collaboration and discussion of various issues discussed in this paper.
This work was partially supported by RFBR grant 04-01-00642, the grant for
support of Scientific
Schools 1578.2003.2, the NWO project 047.017.015, the ANR-05-BLAN-0029-01
project "Geometry and
Integrability in Mathematical Physics" and the Russian Science Support
Foundation.

\section*{Appendix}
\appendix
\setcounter{equation}0
\section{Theta functions and Fay identities
\label{app:Fay}}

In this Appendix we present some definitions and useful formulas for
the Riemann theta-functions, basically taken from the book \cite{Fay}.
One has first to consider
the embedding of the Riemann surface $\Sigma$ into the
$g$-dimensional complex torus ${\bf Jac}$,
the Jacobi variety of $\Sigma$.
This embedding is given, up to an overall
shift in ${\bf Jac}$, by the Abel map
$P \mapsto \bomega(P)= (\omega_1 (P), \ldots , \omega_g (P) )$
where
\be\label{E1s}
\omega_\alpha(P)=\int_{P_0}^{P} d\omega_{\alpha}
\ee
The Riemann theta function $\theta (\bomega) \equiv \theta (\bomega | T)$
is defined by the Fourier series
\be
\label{thetadef}
\theta (\bomega) = \sum_{{\bf n}\in \mathbb{Z}^g} e^{i\pi\bn\cdot T\cdot\bn +
2\pi i \bn\cdot\bomega}
\ee
with the positively definite imaginary part of the period matrix of $\Sigma$
\be
\Im T_{\alpha\beta} \propto \int_\Sigma d\omega_\alpha\wedge d
{\bar\omega}_\beta >0
\ee
The theta function with
a (half-integer) characteristics $\bdelta =
(\bdelta_1, \bdelta_2)$, where $\delta_{\alpha} =
T_{\alpha \beta}\delta_{1,\beta}
+ \delta_{2,\alpha}$ and
$\bdelta_1, \bdelta_2 \in \frac{1}{2}\mathbb{Z}^{g}$ reads
\be
\label{thetad}
\theta_{\bdelta} (\bomega) = e^{i\pi\bdelta_1\cdot T\cdot\bdelta_1 +
2\pi i \bdelta_1\cdot(\bomega+\bdelta_2)}
\theta (\bomega + \bdelta) =
\sum_{{\bf n}\in \mathbb{Z}^g} e^{i\pi(\bn+\bdelta_1)\cdot T\cdot(\bn+\bdelta_1) +
2\pi i (\bn+\bdelta_1)\cdot(\bomega+\bdelta_2)}
\ee
Under shifts by a period of the lattice,
it transforms according to
\be
\label{thetaper}
\theta_\bdelta (\bomega + {\bf e}_\alpha) = e^{2\pi i \delta_{1,\alpha}}
\theta_\bdelta (\bomega)
\\
\theta_\bdelta (\bomega + T_{\alpha\beta}{\bf e}_\beta) =
e^{- 2\pi i \delta_{2,\alpha} - i\pi T_{\alpha\alpha}
-2\pi iW_\alpha}\theta_\bdelta (\bomega)
\ee
The prime form $E(z,\zeta)$ is defined as
\be
\label{primed}
E(P,P' ) = {\theta_\ast (\bomega(P) -\bomega(P'))\over
\sqrt{\sum_\alpha \theta_{\ast,\alpha}d\omega_\alpha(P)}
\sqrt{\sum_\beta \theta_{\ast,\beta}d\omega_\beta(P')}}
\ee
where $\theta_\ast$ is any odd theta
function, i.e., the theta function
with {\em any odd} characteristic
$\bdelta^\ast$ (the characteristics is odd if
$4 \bdelta_1^\ast \cdot \bdelta_2^\ast = {\rm odd}$).
The prime form does not depend on the particular
choice of the odd characteristics.
In the denominator of \rf{primed} we used the notation
\be
\left. \theta_{\ast , \alpha}=
\theta_{\ast , \alpha}(0)=
\frac{\p \theta_{\ast}
(\bomega )}{\p \omega_\alpha}\right |_{\bomega =0}
\ee
for the set of $\theta$-constants.

The data we use in the main text contain also a distinguished
coordinates on a Riemann surface: the holomorphic
co-ordinates $z$ and $\bar z$ on two different sheets of the
Schottky double (see details in \cite{MM1}), and we do not
distinguish, unless it is necessary between the prime form (\ref{primed})
and a {\em function} $E(z,\zeta)\equiv E(z,\zeta)(dz)^{1/2}(d\zeta)^{1/2}$
``normalized" onto the differentials of distinguished co-ordinate.

Let us now list the Fay identities \cite{Fay} used in \cite{KMZ} and above
in the paper.
The basic one is the trisecant identity
(equation (45) from p.~34 of \cite{Fay})
\be
\label{fay1}
\theta (\bomega_1-\bomega_3-\bZ)\ \theta (\bomega_2-\bomega_4-\bZ)\ E(z_1,z_4)E(z_3,z_2)
\\ + \,\,
\theta (\bomega_1-\bomega_4-\bZ)\ \theta (\bomega_2-\bomega_3-\bZ)\ E(z_1,z_3)E(z_2,z_4)
\\ = \,\,
\theta (\bomega_1+\bomega_2-\bomega_3-\bomega_4-\bZ)\ \theta (\bZ)\ E(z_1,z_2)E(z_3,z_4)
\ee
where $\bomega_i \equiv \bomega (z_i)$. This identity
holds for {\em any} four points $z_1,\dots,z_4$ on a complex curve
and {\em any} vector $\bZ \in {\bf Jac}$ in Jacobian.
In the limit
$z_3\to z_4\equiv\infty$ one gets
(formula (38) from p.~25 of \cite{Fay})
\be
\label{fay2}
{\theta ( \int_{\infty}^{z_1}d\bomega +\int_{\infty}^{z_2}d\bomega-\bZ)\
\theta (\bZ)\over
\theta (\int_{\infty}^{z_1}d\bomega-\bZ)\
\theta (\int_{\infty}^{z_2}d\bomega-\bZ)}
{E(z_1,z_2)\over E(z_1,\infty)E(z_2,\infty)} \\ = \,\,
d\Omega^{(z_1 , z_2)}(\infty) + \sum_{\alpha=1}^gd\omega_\alpha(\infty)
\,\, \p_{Z_{\alpha}}
\log\ {\theta (\int_{\infty}^{z_1}d\bomega-\bZ)
\over\theta (\int_{\infty}^{z_2}d\bomega-\bZ)}
\ee
where
\be
\label{abel3}
d\Omega^{(z_1 , z_2)}(\infty)=
d_z\log\ {E(z,z_1)\over E(z,z_2)}
\ee
is the normalized Abelian differential of the third kind
with simple poles at the points with co-ordinates
$z_1$ and $z_2$ and residues $\pm 1$.

Another relation from \cite{Fay} we used above
(see e.g. (29) on p.~20 and (39) on p.~26) is
\be
\label{fay3}
{\theta (\bomega_1-\bomega_2-\bZ)\theta (\bomega_1-\bomega_2 + \bZ)\over
\theta ^2 (\bZ) E^2(z_1,z_2)} = \omega (z_1,z_2) +
\sum_{\alpha , \beta =1}^{g}
\left(\log\theta(\bZ)\right)_{,\alpha\beta}
d\omega_\alpha (z_1)d\omega_\beta (z_2)
\ee
where
\be
\label{A12}
(\log \theta (\bZ))_{, \, \alpha \beta}=
\frac{\p^2 \log \theta (\bZ )}{\p Z_{\alpha}\p Z_{\beta}}
\ee
and
\be
\label{bidif}
\omega (z_1,z_2) = d_{z_1}d_{z_2}\log E(z_1,z_2)
\ee
is the canonical bi-differential of the second kind
with the double pole at $z_1 =z_2$ or the Bergman kernel.

\setcounter{equation}0
\section{Correlation functions in (2,7) and (3,4) minimal string models
\label{app:Ising}}

{\bf $(2,7)$ model}.
The $(p,q)=(2,7)$ model is not much different from the
Yang-Lee case of $(2,5)$ theory considered in sect.~\ref{ss:25}.
The polynomials \rf{polspq} are
\be
\label{pol27}
X= \lambda^2+X_0
\\
Y= \lambda^7 + {7X_0\over 2}\lambda^5+Y_3\lambda^3+Y_1\lambda
\ee
and the calculation of flat times \rf{tP} gives
\be
\label{txy27}
t_1 = {3\over 4}Y_3X_0^2-{105\over 64}X_0^4-Y_1X_0
\\
t_3= {35\over 12}X_0^3-Y_3X_0+{2\over 3}Y_1
\\
t_5 = -{7\over 4}X_0^2+{2\over 5}Y_3
\\
t_7 = 0
\\
t_9 = {2\over 9}
\ee
Again we see, that \rf{txy27} can be easily solved w.r.t. $Y_j$, but
the only coefficient $X_0$ now satisfies
\be
\label{se27}
t_1 = -{35\over 64}X_0^4-{15\over 8}t_5X_0^2-{3\over 2}t_3X_0
\ee
where we put $t_5=0$ for the coefficient at the "boundary" operator \cite{MMS}.

The one-point functions \rf{tP} are given for the $(2,7)$ model by
\be
\label{1p27}
{\d\F\over \d t_1} =
-{7\over 32}X_0^5-{1\over 4}Y_1X_0^2+{1\over 8}Y_3X_0^3
= -{7\over 32}X_0^5 - {5\over 8}X_0^3t_5-{3\over 8}X_0^2t_3
\\
{\d\F\over \d t_3} =
-{1\over 8}Y_1X_0^3-{35\over 512}X_0^6+{3\over 64}Y_3X_0^4
= -{35\over 256}X_0^6-{45\over 128}X_0^4t_5-{3\over 16}X_0^3t_3
\\
{\d\F\over \d t_5} =
-{15\over 512}X_0^7-{5\over 64}Y_1X_0^4+{3\over 128}Y_3X_0^5 =
-{25\over 256}X_0^7-{15\over 64}X_0^5t_5-{15\over 128}X_0^4t_3
\ee
In the r.h.s.'s of \rf{1p27} we already substituted the expressions for
$Y_j$ in terms of times \rf{txy27}, and the rest is to solve \rf{se27} by
expanding in $t_3$ and $t_5$ and substitute result into \rf{1p27}.

The scaling anzatz \rf{scakdv}, \rf{25sca1} now reads
\be
\label{27sca1}
\F = t_5^{9/2}\ {\sf f}\left({t_1\over t_5^2},{t_3\over t_5^{3/2}}\right)
\\
{\d\F\over \d t_1} = t_5^{5/2}{\sf f}^{(1)}, \ \ \ \ \
{\d^2\F\over \d t_1^2} = {X_0\over 2} = t_5^{1/2}{\sf f}^{(11)},\ \ \ldots
\ee
where we have introduced shorten notation for the derivatives over the first argument
of $\f(\t_1,\t_2)$,
and string equation \rf{se27} turns into
\be
\label{27sca}
{\sf t}_1 +{35\over 4}{\sf u}^4 + {15\over 2}{\sf u}^2 + 3\t_2{\sf u} = 0
\ee
for ${\sf u} = {\sf f}^{(11)}$. Solution of
\rf{27sca} (and then of \rf{1p27}) gives for the coefficients of ${\sf f}$
(in "broken phase" with nonvanishing
$\f_{11}$), the following numbers
\be
\label{27fsca}
\f_0 = \frac {180}{2401}\sqrt{-42},\ \
\f_1 = -\frac {6}{49}\sqrt{-42},\ \
\f_2 = \frac {135}{98},\ \
\f_{11}=\frac {1}{7}\sqrt{-42},\ \
\f_{12}=-\frac {9}{7},\ \
\f_{22}=-\frac {18}{49}\sqrt{-42}
\\
\f_{111}= -\frac {1}{90}\sqrt{-42},\ \
\f_{112}=\frac {1}{5},\ \
\f_{122}=\frac {3}{35}\sqrt{-42},\ \
\f_{222}=-\frac {54}{35}
\\
\f_{1111}=-\frac {7}{1620}\sqrt{-42}, \ \
\f_{1112}={14\over 225}, \ \
\f_{1122}= \frac {1}{50}\sqrt{-42},\ \
\f_{1222}= -\frac {6}{25},\ \
\f_{2222}=-\frac{9}{175}\sqrt{-42}
\\
\f_{11111}=-\frac{343}{81000}\sqrt{-42},\ \
\f_{11112}=\frac{196}{3375},\ \
\f_{11122}=\frac{49}{2700}\sqrt{-42},\ \
\f_{11222}-\frac{28}{125},\ \
\ldots
\ee
and one can easily compute the corresponding invariant ratios, for example
\be
{\f_{1111}\f_{22}\over \f_{112}^2} = -{5\over 3}, \ \
{\f_{2222}\f_{11}\over \f_{122}^2} = -1, \ \
{\f_{1111}\f_{11}\over \f_{111}^2} = -5, \ \
{\f_{2222}\f_{22}\over \f_{222}^2} = -{1\over 3},
\ \ \ldots
\ee
{\bf Ising model}. Now let us turn to the "two-matrix model" Ising point $(p,q)=(3,4)$.
Here the calculation of residues \rf{tP} (upon vanishing $t_3=0$ and
$t_4=0$) gives rise to
\be
\label{coef34}
Y_2 = {4\over 3}X_1 + {5\over 3}t_5
\\
Y_1= {4\over 3}X_0 
\\
Y_0= {2\over 9}X_1^2 + {10\over 9}X_1t_5
\ee
while $X_0$ and $X_1$ satisfy
\be
\label{seis}
t_1=-{2\over 3}X_0^2+{4\over 27}X_1^3 +{5\over 9}t_5X_1^2 
\\
t_2=-{2\over 3}X_0X_1-{5\over 3}t_5X_0 
\ee
Upon solving the second equation for $X_0$, this system turns into an algebraic
Kazakov-Boulatov equation \cite{KB} for a single function $X_1$
\be
\label{kbeq}
t_1=-{6t_2^2\over (2X_1+5t_5)^2}+{4\over 27}X_1^3 +{5\over 9}t_5X_1^2
\ee
which contains all information about singularities of spherical partition function
for arbitrary values of magnetic field $t_2$ and fermion mass $t_5$ \cite{Zamolun}.

The one-point functions \rf{tP} are here given by
\be
\label{ising1}
{\d\F\over \d t_1} = {1\over 27}X_1^4
+{10\over 81}t_5X_1^3 -{4\over 9}X_1X_0^2 - {5\over 9}t_5X_0^2
\\
{\d\F\over \d t_2} = {4\over 27}X_1^3X_0+{10\over 27}t_5X_1^2X_0-{8\over 27}X_0^3
\\
{\d\F\over \d t_5} = {40\over 243}X_1^3X_0^2-{10\over 2187}X_1^6
+{25\over 81}t_5X_1^2X_0^2-{10\over 729}t_5X_1^5-{5\over 27}X_0^4
\ee
Again, using \rf{sysi}, or
differentiating the first equation of \rf{ising1} and using \rf{seis} one gets
\be
\label{uis}
{\d^2\F\over \d t_1^2} = {X_1\over 3}, \ \ \ {\d^2\F\over \d t_1\d t_2} = {2X_0\over 3},
\ \ \ldots
\ee
The unitary scaling anzatz \rf{scaF} suggests that the rescaled function
$f$ depends only upon dimensionless quantities $\tau_\sigma=t_2/ t_1^{5/6}$ and
$\tau_\epsilon=t_5/t_1^{1/3}$:
\be
\label{scaisi}
\F = t_1^{7/3} f\left( \tau_\sigma,\tau_\epsilon\right)
\ee
where two arguments of $f$ naively correspond to the "Liouville-dressed"
Ising operators $\sigma$ and $\epsilon$ respectively.

Note also, that for $t_2=0$ the second equation of \rf{seis} has the only reasonable
solution $X_0=0$, while the first one turns into
\be
\label{isred}
t_1={4\over 27}X_1^3 +{5\over 9}t_5X_1^2
\ee
which coincides with the perturbation of the Yang-Lee $(2,5)$ model by quadratic term
(see \rf{gpse}; note also that one has to identify ${X_1\over 3}$ from \rf{isred} with
${X_0\over 2}$ from \rf{gpse}, cf. with the formulas \rf{252d} and \rf{uis}).
For $t_2=0$ the function \rf{scaisi} becomes a function of only the second argument
\be
\label{scaisir}
\F = t_1^{7/3} \hat f\left({t_5\over t_1^{1/3}}\right) =
\\ =
t_1^{7/3}f_0 + t_1^2t_5f_\epsilon +
+ \ha t_1^{5/3}t_5^2f_{\epsilon\epsilon} +
{1\over 6}t_1^{4/3}t_5^3f_{\epsilon\epsilon\epsilon}
+ {1\over 24}t_1t_5^4f_{\epsilon\epsilon\epsilon\epsilon}
+ {1\over 120}t_1^{2/3}t_5^5f_{\epsilon\epsilon\epsilon\epsilon\epsilon}
+ \ldots
\ee
Using \rf{isred} one gets for \rf{scaisir}
\be
\label{ecof}
f_0={9\over 56}2^{1/3}, \ \ \
f_\epsilon = -{5\over 24}, \ \ \
f_{\epsilon\epsilon} = {5\over 32}2^{2/3}, \ \ \
f_{\epsilon\epsilon\epsilon} = -{125\over 576}2^{1/3},\ \ \
f_{\epsilon\epsilon\epsilon\epsilon\epsilon} = -{3125\over 20736}2^{2/3}
\ee
while the coefficient $f_{\epsilon\epsilon\epsilon\epsilon}$ remains undetermined from
\rf{isred}. To find it one can use the first line of \rf{ising1} (at $X_0=0$), which
gives
\be
\label{f4}
f_{\epsilon\epsilon\epsilon\epsilon} = {625\over 96}
\ee
However, under reparameterization in the space of couplings (see e.g. \cite{MSS})
\be
X_1 \rightarrow X_1 - {5\over 12}t_5
\\
t_1 \rightarrow t_1 + {125\over 216}t_5^3
\ee
the reduced string equation \rf{isred} acquires the form of (analytically continued)
string equation \rf{gpse} for the Yang-Lee model. Therefore, one can use further the
scaling anzatz \rf{25sca} with $t_3=t_3^{\rm YL}$ of the $(2,5)$ critical point
substituted by the {\em square} of the $t_5=t_5^{\rm Ising}$ of the (reduced) Ising model,
so that the expansion \rf{gpis} would give the $\langle\epsilon^{2n}\rangle$
correlators of the gravitationally dressed $(3,4)$ Ising model.

\setcounter{equation}0
\section{Residues and equations for the logarithmic potential
\label{app:Miwa}}

Let us present here some useful explicit formulas for the two-matrix model with
logarithmic potential \rf{jpot}, in addition to sect.~\ref{ss:gp}.

The generating differential $\tilde z dz$ \rf{gendiff} possesses four nontrivial
residues at the points
\be
\label{points}
q_1 = \left.z\right|_{w=0}= v-{u\over s} = -{\nu\over a}
\\
q_2 = \left.z\right|_{w=1/s}= v+{r\over s}+{us\over 1-s^2} = a
\\
\infty_- = \left.z\right|_{w=\infty}
\\
\infty_+ = \left.z\right|_{w=s}
\ee
(one finds for the parameter \rf{lam} that $\Delta=q_2-q_1$), which are
\be
\label{res}
{1\over 2\pi i}\ \res_{q_1} {\tilde z}dz = - {1\over 2\pi i}\ \res_{\infty_-} {\tilde z}dz =
r^2 - {ru\over s^2}
\\
{1\over 2\pi i}\ \res_{q_2} {\tilde z}dz = - {1\over 2\pi i}\ \res_{\infty_+} {\tilde z}dz =
{u^2\over (1-s^2)^2} -{ru\over s^2}
\ee
Taking derivative of the generating differential \rf{gendiff}
w.r.t. the period $S$ \rf{per} at fixed $t_0$ \rf{t0conf} and introduced
by \rf{delta} extra parameter $\delta$,
one gets the canonical holomorphic differential on \rf{jcur}, \rf{jcurd}
\be
d\omega = {\d\over\d S}\ \tilde z dz =
- {\d h\over\d S}{dz\over \sqrt{R(z)}}
\ee
provided by normalization condition
\be
\label{halpha}
{\d h\over\d  S}\oint_A {dz\over \sqrt{R(z)}} = - 1
\ee
Similarly,
\be
{\d h\over\d t_0} \oint_A {dz\over\sqrt{R}} =
\oint_A dz\left({1\over 2\left(z+{\nu\over a}\right)}{P\over\sqrt{R}}+
{az\over\sqrt{R}}\right)
\ee
and this completes the set of conditions, fixing all parameters in the
logarithmic model of sect.~\ref{ss:gp}.

In addition to the main cubic equation (\ref{cueqX}),
one can write a similar equation for the square of
conformal radius (\ref{ru})
\be
\label{rho}
\rho \equiv r^2 = {t_0\over 4}
\left({{\cal X}\over t_0/\Delta^2} + {t_0/\Delta^2\over {\cal X}} + 2\right)
\ee
which satisfies
\be
\label{cueqrho}
\rho^3-\ha\left({\Delta^2\over 4}+{7t_0\over 2}-\nu\right)\rho^2+
{t_0\over 2}\left({\Delta^2\over 4}+{3t_0\over 2}+{t_0\nu\over 2\Delta^2}
-{t_0^2\over 4\Delta^2}-\nu\right)\rho +{t_0^2\over
8\Delta^2}(t_0-\nu)^2=0
\ee
The critical points of this cubic curve are
at $t_0=0$, $\rho=0$; $t_0={\Delta^2\over 4}\left(1\pm
3\sqrt{1-{16\nu\over 9\Delta^2}}\right)$.
In the same way it is easy to show that the squared length of the cut
\be
\Xi={4ru\over\Delta^2} = \left({\xi_+-\xi_-\over 2\Delta}\right)^2 =
{t_0^2\over\Delta^4} +  \left({1\over 2} + {2\nu\over\Delta^2} -
{t_0\over\Delta^2}\right){\cal X} -{t_0^2\over 2\Delta^4}{1\over {\cal X}}
\ee
also satisfies the cubic equation
\be
\Xi^3 + \left({(t_0-2\nu)^2\over \Delta^4}-{3t_0\over\Delta^2}
 -{1\over 4}\right)\Xi^2 + {t_0\over \Delta^2}\left({1\over 2}+
{7t_0\over 2\Delta^2}-{4\over \Delta^4}(t_0-\nu)(t_0-2\nu)\right)
\Xi -
\\
- {t_0^2\over\Delta^4}\left({1\over 4}+
{3\over 2}{t_0\over\Delta^2}-{15t_0^2\over 4\Delta^4}+
{4\nu\over\Delta^4}(3t_0-2\nu)+{2\over\Delta^6}(t_0-2\nu)^3\right) = 0
\ee

\end{document}